\documentclass[reprint,twocolumn,aps,prd,superscriptaddress,floatfix,amssymb]{revtex4-1}
\usepackage{amsmath}
\usepackage{graphicx}  % needed for figures
\usepackage{dcolumn}   % needed for some tables
\usepackage{bm}        % for math
\usepackage{amssymb}   % for math
\usepackage[mathlines]{lineno}% Enable numbering of text and display math
\usepackage{subfigure}
\usepackage{color}
\usepackage{url}
\usepackage{cancel}
\usepackage{booktabs}
\usepackage{blindtext}

\newcommand{\aap}{{A\&A}}

\newcommand{\aj}{{Astro.~Journal}}
\newcommand{\procspie}{{Proc.~Astron.~Soc.~Pac.}}
\newcommand{\mnras}{{Mon.~Not.~R.~Astron.~Soc.}}
\newcommand{\jcap}{{J.~Cosmol.~Astropart.~Phys.}}
\newcommand{\apjl}{{Astrophys.~J.~Lett.}}
\newcommand{\pasj}{{Publ Astron Soc Jpn }}
\newcommand{\pasp}{{PASP}}

\begin{document}

\widetext

\title{Cosmology with Galaxy Cluster Phase Spaces}

\author{Alejo Stark}
\affiliation{Department of Astronomy, University of Michigan, Ann Arbor, MI 48109 USA}
\author{Christopher J. Miller} 
\affiliation{Department of Astronomy, University of Michigan, Ann Arbor, MI 48109 USA}
\affiliation{Department of Physics, University of Michigan, Ann Arbor, MI 48109 USA}
\author{Dragan Huterer}
\affiliation{Department of Physics, University of Michigan, Ann Arbor, MI 48109 USA}

\date{\today}

\begin{abstract}
We present a novel approach to constrain accelerating cosmologies with galaxy cluster phase spaces. With the Fisher matrix formalism we forecast constraints on the cosmological parameters that describe the cosmological expansion history. We find that our probe has the potential of providing constraints comparable to, or even stronger than, those from other cosmological probes. More specifically, with 1000 (100) clusters uniformly distributed in the redshift range $ 0 \leq z \leq 0.8$, after applying a conservative $80\%$ mass scatter prior on each cluster and marginalizing over all other parameters, we forecast $1\sigma$ constraints on the dark energy equation of state $w$ and matter density parameter $\Omega_M$ of $\sigma_w = 0.138 (0.431)$ and $\sigma_{\Omega_M} =  0.007 (0.025)$ in a flat universe. Assuming 40\% mass scatter and adding a prior on the Hubble constant we can achieve a constraint on the CPL parametrization of the dark energy equation of state parameters $w_0$ and $w_a$ with 100 clusters in the same redshift range: $\sigma_{w_0} =  0.191 $ and $\sigma_{w_a} =  2.712$. Dropping the assumption of flatness and assuming $w=-1$ we also attain competitive constraints on the matter  and dark energy density parameters:  $\sigma_{\Omega_M} =  0.101$ and $\sigma_{\Omega_{\Lambda}} =   0.197$ for 100 clusters uniformly distributed in the range $ 0 \leq z \leq 0.8$ after applying a prior on the Hubble constant. We also discuss various observational strategies for tightening constraints in both the near and far future. \end{abstract}

\pacs{}
\maketitle

\section{Introduction}
The relation between the dynamics of clustered galaxies and the totality of the observable universe has long been a fruitful site of investigation for physical cosmology. One good example is Zwicky's seminal investigation of the dark matter, which crystallized through an analysis of the dynamics of galaxies in the Coma cluster (then referred to as "clusters of nebulae")  \cite{zwicky}. Three decades ago, also utilizing the Coma cluster as a cosmological laboratory, Ref. \cite{shectman} demonstrated that the universe's matter energy density was sub-critical -- thereby pointing to the existence of some other unknown substance and demonstrating the capacity for the  infall regions of galaxy clusters to be mobilized as a powerful cosmological probe. Just a few years after this, Ref. \cite{regosGeller} used the Coma cluster in conjunction with three other nearby clusters to argue for the existence of ``caustics'' in the redshift-distance space of galaxy clusters. These ``caustics'' trace an identifiable edge that would later be demonstrated to be representative of the radial escape velocity profile of galaxy clusters \cite{diaferiogeller1997}. From this point on, the caustics would be used solely to generate mass profiles of galaxy clusters via the so-called caustic technique \cite{diaferio1999,geller2013}. These dynamical cluster masses have been used to constrain cosmology through the cluster mass function \cite{rines2008}. However, the capacity of caustic edges themselves to constrain cosmological parameters has not been pursued since the work of Ref. \cite{regoes1996}.

Following Refs.  \cite{shectman, regosGeller, regoes1996}
but also deviating significantly from their approach, Ref. \cite{starkApJ} demonstrated the necessity to include a cosmological term to describe the escape velocity profile of galaxy clusters as inferred from their projected phase spaces. In particular, Ref. \cite{starkApJ} presented an analytic model based on the Poisson equation that can reproduce the projected escape velocity profiles of galaxy clusters as measured from their phase spaces. The analytical escape velocity profile prediction requires a known mass profile and a known velocity anisotropy profile $\beta(r)$. Given these and the cosmological parameters, the analytical escape velocity edge has been shown to match expectations to high precision and accuracy using N-body simulations \citep{miller2016}. 

If both weak lensing mass estimates and a measurement of $\beta$ can be inferred for a galaxy cluster, we can turn this around and through analytic theory constrain cosmology by measuring edges of clusters through their phase spaces (see Refs.  \cite{diaferio1997, serra, geller2013,lemze2009,miller2016} for the various methods utilized to estimate the escape velocity "edge" from the phase space of galaxy clusters).

However, while Ref. \cite{starkApJ}  demonstrated the necessity to include a cosmological and redshift-evolving term in the escape velocity profile of clusters, it did not quantify the precision with which one can constrain cosmological parameters with this observable. This is the task that this paper takes on.

Our paper is organized as follows: in Section II we describe the theoretical observable we work with -- the redshift-evolving and cosmology dependent escape velocity profile of galaxy clusters. In Section III we detail how we use the Fisher matrix formalism to quantify how well this observable can constrain cosmology given current and future systematic errors. We present the results of this analysis in Section IV. In Section V we present observational strategies that may be utilized to optimize cosmological constraints. In Section VI we discuss our observable in relation to other probes and speculate as to how we may improve constraints in the future through joint likelihood analyses. We conclude the paper and provide ways of extending our work in Section VII. The Appendix details the construction of our Fisher matrix.
%%%%%%%%%%%%%%%%
%%%%%%%%%%%%%%%%
%%% observable   %%%
%%%%%%%%%%%%%%%%
%%%%%%%%%%%%%%%%
\section{theoretical observable}

The theoretical observable quantity we work with throughout this paper is the projected escape velocity radial profile of galaxy clusters.  Observationally, the escape velocity profile of a cluster is inferred from the phase space ($v_{los}$ vs $r$ space) of the cluster. More specifically, the line of sight galaxy velocity ($v_{los}$) vs. physical distance ($r$) space is constructed by measuring the redshifts of cluster galaxies ($z$) of a cluster at a redshift $z_c$ and then converting them to velocities via,
\begin{equation} 
v_{los}(r) =  c\frac{(z - z_c)}{{(1+z_c)} },
\end{equation} 
where $c$ is the speed of light. The physical distance from the cluster's center ($r$) is inferred from the angular diameter distance ($d_A$) and the measured angular separation ($\theta$), 

\begin{equation} 
r = d_A(z) \theta = \Bigg[ \frac{1}{1+z}\frac{c H_0^{-1}}{\sqrt{\Omega_K}} \text{sin} \bigg( \sqrt{\Omega_K}  \int_0^z \frac{dz'}{E(z')}  \bigg)\Bigg]\theta. \label{eq:Da}
\end{equation}
$H_0$ is the Hubble constant and the redshift evolving function $E(z)$ is detailed in Eq.~(\ref{eq:hubbleparam}) below. Note that sin($x$) (the non-flat closed universe case) becomes $x$ for the flat universe case. Also detailed below is the parameter $\Omega_K$ which quantifies the openness or closedness of the universe. The escape velocity profile can then be inferred from this phase space through various techniques (see Refs. \cite{diaferio1997, serra, geller2013,lemze2009,miller2016}). 

This phase space inferred escape velocity profile can be modeled with a function of the mass distribution of a specific cluster as specified by its gravitational potential (in our case we use the Einasto profile with three free parameters: $\alpha, r_{-2}, \rho_{-2}$; see Eq.~(\ref{eq:einasto_pot}) below), and the anisotropy parameter ($\beta$) of that specific cluster (which allows us to take into account projection effects). As mentioned in the introduction, the profile is also a function of redshift $z$ and cosmology ($\Omega_M, h$, etc.).  The escape velocity radial profile is therefore given by a function of these cosmology and cluster parameters combined,
\begin{equation} 
v_{esc}(r,z,\Omega_M, h, \dots, \beta, \alpha,  r_{-2},\rho_{-2}).
\end{equation}
We note that while we utilize only Einasto density profiles in this paper, in principle any parametrized mass profile can be used in our framework, as long as the parametrization is a density-potential Poisson pair (see e.g., Ref. \cite{miller2016}). 

While the mass dependence and projection effects have long been considered in studies of this observable (see for instance Ref. \cite{diaferio1997}), only recently has the cosmological dependence of the escape velocity profile been considered. More specifically, the cosmological dependence of the escape velocity has been studied in relation to both observational data and simulations of standard general relativistic cosmology as well as modified theories of gravity \cite{starkApJ,miller2016,starkmiller}. These investigations  found  the need to include a cosmological term in order to reproduce numerical results as well as observational data.

Qualitatively, it should be unsurprising that the cosmological background within which a cosmic structure is embedded at a given epoch will create the conditions for its evolution and development. For instance, whether a gravitationally bound structure today (say, a galaxy group or galaxy cluster) can become unbound at late times  in an accelerating universe is a function of both the curvature of space and the mass-energy content of the background the particular structure is embedded in \cite{busha,behroozi}. It is therefore clear that the theoretical escape velocity profile must take into account the cosmological background.

More specifically, the cosmological dependence in the escape velocity profile in Refs. \cite{starkApJ,miller2016,starkmiller} comes in through the "equivalence radius." In an accelerating universe, the radius out to which one has escaped a cluster's potential is a function of cosmology. This minimal radius required to escape, termed the equivalence radius ($r_{eq}$), decreases in an accelerating universe (see Ref. \cite{starkApJ} and references therein). The projected escape velocity profile is then given by,

\begin{equation} 
v_{esc}(r,z) = \sqrt{\frac{1}{g(\beta)}\bigg[-2 \big(\Psi(r) - \Psi(r_{eq}) \big) - qH^2 \big(r^2 - r_{eq}^2 \big)\bigg]}.\label{eq:vesc}
\end{equation}
We note that, as expected, at $r = r_{eq}$ the escape velocity is nil. The function of the equivalence radius then is to normalize the escape velocity at that point. Note that this equation is derived by integrating the acceleration equation of the effective potential that takes into account both the "negative" acceleration due to the mass of the cluster and the "positive" acceleration of the background at late-times. We therefore obviate the "negative" acceleration due to ram pressure on galaxies given that it is negligible when compared to averaged gravitational effects \cite{Faltenbacher}. The "equivalent radius," then, is named for the condition that it sets: it is the point at which the negative (inward) acceleration due to the pull of the cluster and the positive (outward) acceleration due to the acceleration of the universe balance each other. For a derivation of  Eq.~(\ref{eq:vesc}) see Refs. \cite{starkApJ,miller2016}. We note also that the physical  distance $r$ in Eq.~(\ref{eq:vesc}) is cosmology dependent via Eq.~(\ref{eq:Da}).

Lastly, as detailed in Ref. \cite{starkApJ},  Eq.~(\ref{eq:vesc}) emerges from the acceleration equation derived by Ref. \cite{Nandra2012} in the context of a flat universe. However, in the cosmological regime we work with, the acceleration equation for a spatially flat and non-flat universe converge. See sections 5.3.1, 5.3.2, and 5.3.3 of Ref. \cite{Nandra2012} and section 2.1 of Ref. \cite{starkApJ}.

We now consider the two main components of the projected escape velocity profile function (Eq.~(\ref{eq:vesc})): cluster parameters (projection effects and mass profile information)  and cosmological parameters.

\subsection{Cluster parameters}
The two components of the cluster parameter set are the mass profile parameters and the anisotropy parameter $\beta$ encapsulated by the function $g(\beta).$ 

The anisotropy parameter $\beta$ is given by $\beta(r) = 1- \sigma_t^2 / \sigma_r^2.$ Where $\sigma_t$ is a function of the azimuthal and tangential velocity dispersions and $\sigma_r$ is the radial velocity dispersion. See Section 5.1 of Ref. \cite{starkApJ} for details. Note that while the anisotropy profile of the cluster is actually a function of radius ($\beta(r)$), the observed and simulated profile is nearly flat within 0.3 - 1 virial radii \cite{lemze2013,serra,munari}. Therefore, in what follows we only consider the escape velocity profile within this radial range so that we can reduce the anisotropy profile to a single value for a given cluster.  Once $\beta$ for a given cluster is inferred we can then use the function $g(\beta)$ to project our escape velocity profile (see Refs. \cite{diaferio1997,starkApJ}). In particular this function is defined geometrically and given by,
\begin{equation}        
g(\beta) = \frac{3-2\beta}{1-\beta}.
\end{equation}
As can be implied from this equation, the effect of $g(\beta)$ on the escape velocity profile within the radial range we consider below is to suppress the profile by a constant value when compared to the 3 dimensional case ($g(\beta) = 1$). Quantitatively, the limits of this function are set by the limiting cases of the anisotropy parameter $\beta$: radial 
infall ($\beta$ = 1), circular motion ($\beta = - \infty$) and isotropy ($\beta = 0$). For the radial range we are considering, on average, $g(\beta) \sim 3.3$. This entails that, on average, within the radial range considered below, non-projected escape velocity profiles are $ \sqrt{g(\beta)} \sim 1.8 $ times higher than projected profiles. As a rule of thumb, the more radial the velocity anisotropy of a cluster, the more suppressed the escape velocity profile will be. For a more thorough quantitative analysis of these projection effects see Refs. \cite{starkApJ,serra}.

The other component of the cluster parameters that make up our observable is the mass profile of a given cluster. Information about the mass profile of the cluster comes through the gravitational potential $\Psi(r)$. Following \cite{starkApJ,miller2016} we pick the Einasto representation of the potential given its capacity to trace the mass distribution of galaxy clusters beyond the virial radius. The gravitational potential $\Psi(r)$, then, is a function of three free parameters:  the shape parameter $\alpha$, the radius where the logarithmic slope of the density profile is equal to $-2$ ($r_{-2}$) and  the density at $r_{-2}$ ($\rho_{-2}$).
 As calculated via the Poisson equation  \cite{retana}, the potential as inferred from an Einasto density field is,

\begin{equation}        
\Psi(r) = -\frac{{ GM}}{r} \Bigg{[} 1 - \frac{\Gamma\big{(}3/\alpha,s^{\alpha}\big{)}}{\Gamma(3/\alpha)} + s \frac{\Gamma\big{(}2/\alpha,s^{\alpha}\big{)}}{\Gamma(3/\alpha)}\Bigg{]}. \label{eq:einasto_pot}        
\end{equation}        
$\Gamma(a,x)$ is the upper incomplete gamma function. We are also utilizing the unitless scale radius $s$ given by,
\begin{equation}
s = \frac{ r}{ r_{-2}} \bigg(\frac{2}{\alpha}\bigg)^{1/\alpha}.
\end{equation}        
And we re-write the total mass ($M$) as,
\begin{equation}
M = 4 \pi \rho_{-2}  r_{-2}^3  F(\alpha).
\end{equation}        
In this last equation we have defined the function of $\alpha$, 
\begin{equation}
F(\alpha) = \Gamma(3/\alpha) \bigg( \frac{e^2}{8} \alpha^{3-\alpha} \bigg)^{1/\alpha}.
\end{equation}        
Note that in general we follow the definitions and specifications used in Refs.  \cite{miller2016,starkApJ}.
We treat the mass profiles and $\beta$ as observable quantities with uncertainties. In simulations, $\beta$ shows no significant evolution as a function of redshift in the range we are interested in  (ie., $ 0 \leq z \leq 0.8$, see Fig. 2 in Ref. \cite{iannuzzi}) and is inferred independently of cosmological assumptions \cite{lokasabell,benatov,lemze2009,wojtak,macs1206}. This is not the case for the weak lensing mass profile-inferred parameters (namely $\alpha, r_{-2},$ and $\rho_{-2}$). We analyze how systematic errors introduced by uncertainties in cosmological parameters affect weak lensing mass errors in the section IIIC below as well as in the Appendix.

\subsection{Cosmological parameters}
Having defined the first set of parameters of the escape velocity profile related to its mass content ($\alpha, r_{-2}, \rho_{-2}$) and projection effects ($\beta$), let us now focus on the redshift-evolving and cosmology-dependent terms in Eq.~(\ref{eq:vesc}), namely, the Hubble parameter $H$, the deceleration parameter $q$, and the equivalence radius $r_{eq}$.  We describe these terms below.

Given that we work in a regime where the radiation energy density is negligible ($z \leq 0.8$) the  Hubble parameter is given by,
\begin{multline} 
H^2 = H_0^2 E^2(z)
= H_0^2 \bigg[ \Omega_M (1+z)^3 + \Omega_k (1+z)^2 + \\ \Omega_{DE} \exp\bigg\{ 
 3 \int_0^{z} d\ln (1+x) [1+w(x)] \bigg\} \bigg],
\label{eq:hubbleparam}
\end{multline}
where $\Omega_M$ and $\Omega_{DE}$ are energy densities in matter and dark energy relative to critical, $w(z)$ is the time-varying equation of state of dark energy, and the spatial curvature density parameter is $\Omega_k \equiv 1 - \Omega_M - \Omega_{DE}$. Note that throughout this paper we work with the scaled Hubble constant $h$ defined via $H_0 = 100  h$.  

Once we have the Hubble parameter as a function of redshift we can derive the corresponding deceleration parameter, $q \equiv - \ddot{a}a/ \dot{a}^2$, where $a$ is the scale factor, as a function of redshift via,
\begin{equation}
q = \frac{(1+z)}{H} \frac{dH}{dz} -1.
\label{eq:qfunction}
\end{equation}

Lastly, the equivalent radius ($r_{eq}$) is the physical distance at a given cosmic epoch, for a given cosmology, where the inward pull of gravity balances the outward pull of cosmic acceleration. It is given by \cite{behroozi},

\begin{equation} 
r_{eq} =  \bigg(\frac{GM}{-qH^2}\bigg)^{1/3}.
\label{eq:req}
\end{equation}

This radius is what sets the normalization of the gravitational potential of our galaxy cluster, $\Psi(r_{eq})$ in Eq.~(\ref{eq:vesc}) \cite{miller2016, starkApJ}. That is, beyond this equivalence radius,  the effective potential of the cluster described by Eq.~(\ref{eq:vesc}) ($-2\phi = v_{esc}^2(r,z) $) is normalized to 0 (See Ref. \cite{starkApJ} for a derivation). Note also that the distance at which this balance of forces occurs, that is, between cosmic acceleration and the cluster's gravitational pull, is far away enough from the cluster center that the cluster can be represented as a point mass. Lastly, this quantity can be thought of as a proxy for how sensitive $v_{esc}(r,z)$ is to cosmology. In lieu of taking analytic derivatives of  Eq.~(\ref{eq:vesc}) one can study the analytic derivatives of $r_{eq}$. We have confirmed this by looking at numerical derivatives of $v_{esc}(r,z)$ and compared them to the analytic derivatives of $r_{eq}$ for the same cosmology, finding good agreement. 

\begin{figure}
  \centering
  \includegraphics[width=1\linewidth]{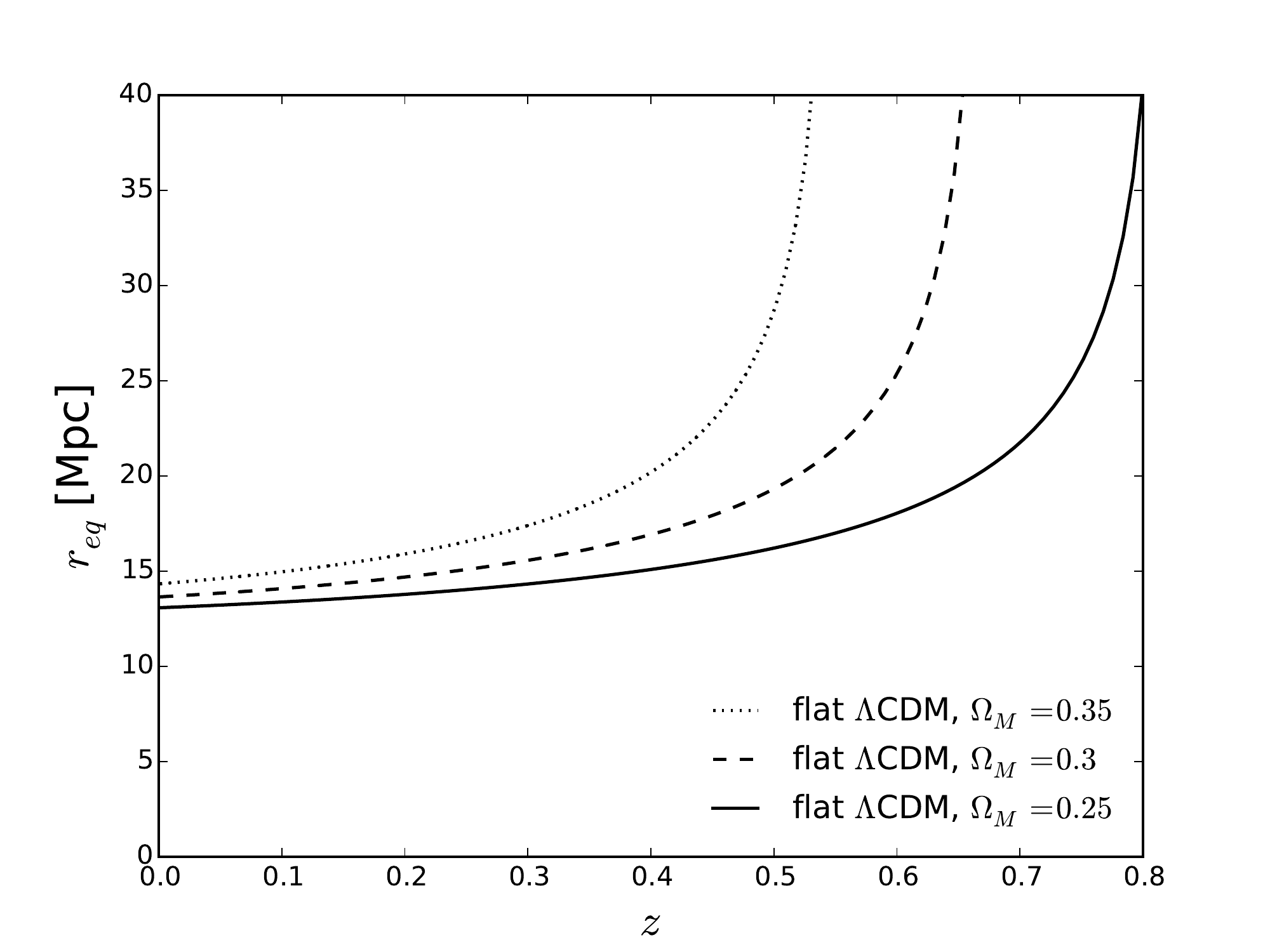}
  \caption{Behavior of the equivalent radius ($r_{eq}$) as a function of redshift ($z$) for a cluster with mass $M_{200} = 4 \times 10^{14} M_{\odot}$  for three values of $\Omega_M$ in a flat $\Lambda$CDM universe and $h=0.7$. At the transition redshift for each given cosmology, the radius shoots up to infinity. Note how $r_{eq}$ becomes more and more sensitive to $\Omega_M$ at higher $z$.}\label{fig1}
\end{figure}

In Fig.~\ref{fig1} we show that, for a fixed cluster mass the equivalence radius is sensitive to both cosmology and redshift. More specifically, we pick a mass of $M_{200} = 4 \times 10^{14} M_{\odot}$, where $M_{200}$ is defined as the mass enclosed by a sphere with an average density equivalent to 200 times the mean density of the universe. Note that as the cluster gets closer to the acceleration transition redshift ($q(z)=0$) the equivalence radius shoots up to infinity. We consider the effects of this behavior on our observable below.

\subsubsection{$v_{esc}(r,z)$ in an accelerating universe ($q < 0$)}
Within the $q < 0$ regime, the escape velocity profile is described by Eq.~(\ref{eq:vesc}). As shown by Ref. \cite{starkApJ}, in an accelerating universe the  effect on the observable generated by changing the various cosmological parameters is to modify both the amplitude and the shape of the escape velocity profile. For instance, in a flat universe, a larger dark energy density makes it easier to escape a galaxy cluster, while a larger dark matter component makes it harder to escape the cluster. This is illustrated in Fig.~\ref{fig1}. Using the equivalence radius as a proxy to gauge how $v_{esc}(r,z)$ changes, we see that the equivalence radius blows up at the given transition redshift for that cosmology (i.e. $z$ that yields $q(z)=0$).

\begin{figure}
  \centering
  \includegraphics[width=1\linewidth]{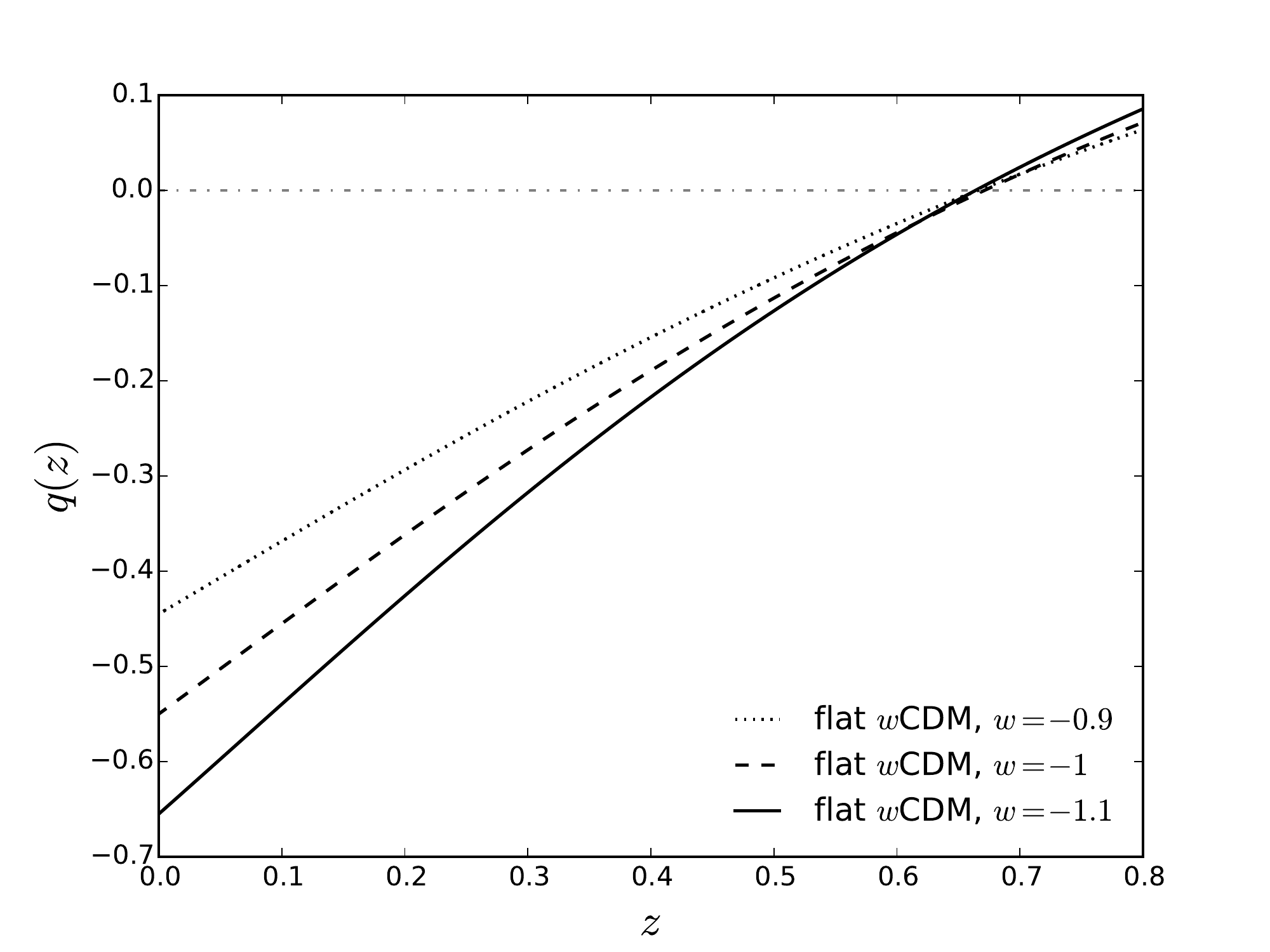}
  \caption{Redshift evolution of the deceleration parameter in a flat universe with fixed $\Omega_M =0.3$ and $h=0.7$. Notice the divergence of values of $q$ at low redshifts when the equation of state parameter is varied.}\label{fig2}
\end{figure}

\begin{figure}
  \centering
  \includegraphics[width=1\linewidth]{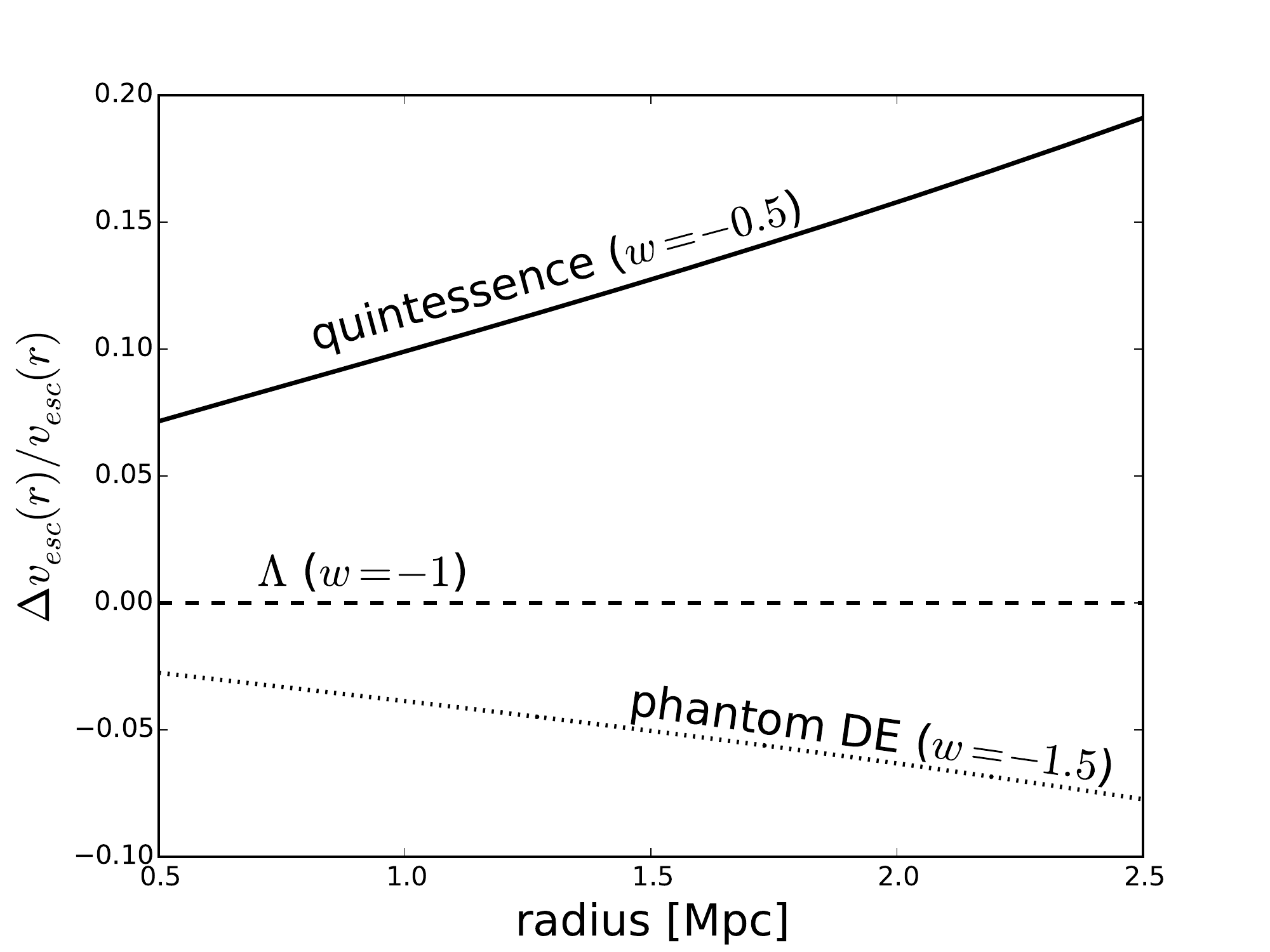}
  \caption{Using a single cluster with mass $M_{200} = 4 \times 10^{14} M_{\odot}$ at $z=0$, we show the fractional difference between the escape velocity profile of a flat $w$CDM universe with dark energy equation of state $w=-1$ and two other dark energy models (quintessence in the solid line and phantom dark energy in the dotted line). More specifically, $\Delta v_{esc}(r,z) /v_{esc}(r,z)= [ v_{esc}(w) / v_{esc}(w=-1)  ] -1 .$  Quintessence therefore acts similarly to increasing the dark matter density. That is, it increases the escape velocity profile relative to the $ \Lambda$ case (see Fig. 1 of Ref. \cite{starkApJ}). On the other hand, the phantom dark energy suppresses the escape velocity profile relative to the $\Lambda$ case. Lastly, we highlight that the fractional difference increases with radius in both cases.}\label{fig3}
\end{figure}

It is important to emphasize that, in the regime where $q(z) < 0$, $v_{esc}(r,z)$ is a direct measure of both expansion and acceleration, since $qH^2$ is a function proportional to both $H(z)$ and $dH(z)/dz$. This makes our observable a powerful probe of cosmology. For example, note the sensitivity of the deceleration parameter to the dark energy equation of state $w$ as shown in Fig.~\ref{fig2}. In particular, the variation of $q(z)$ with respect to $w$ increases at lower redshifts. Fig.~\ref{fig3} also demonstrates the sensitivity of the observable to $w$  by showing the fractional difference between the escape velocity for a flat $\Lambda$CDM universe at $z=0$  and two dark energy models, quintessence-like dark energy (solid line) \cite{quintessence} and phantom dark energy (dotted line) \cite{caldwell}. Note the $\sim$10\% level differences between these models and $\Lambda$ (dashed line).  A quintessence-like dark energy, then, would increase the escape velocity profile, whereas a phantom dark energy would decrease the escape velocity (both relative to $\Lambda$CDM). 

\subsubsection{ $v_{esc}(r,z)$ at the transition redshift ($q = 0$)}
As implied in the previous subsection,  as we approach $q=0$ the equivalence radius blows up and Eq.~(\ref{eq:vesc}) is reduced to the gravitational potential described by the Einasto gravitational potential $\Psi(r)$.  More explicitly, 
\begin{equation} \label{eq:v_esc_limit}
\lim_{q\to0} v_{esc}(r,z) =  \sqrt{\frac{1}{g(\beta)}\bigg[-2 \Psi(r)\bigg]}.
\end{equation}
One immediate consequence we derive from this behavior is that the only cosmological dependence we get beyond this point is through $r$ (see Eq.~(\ref{eq:Da})).

We now investigate what happens to our observable beyond the transition redshift or, equivalently, what happens to our observable with combinations of cosmological parameters that yield a decelerating universe (i.e. $q > 0$).

\subsubsection{ $v_{esc}(r,z)$ beyond the transition redshift ($q > 0$)}

In the regime beyond the transition redshift, or when $q >0$, it makes no sense to speak of an equivalence radius. Recall that the equivalence radius is defined in the context of a balance of forces. Given that there is no balance of forces between the mass-induced pull of gravity and the repulsive acceleration as there is in the case where $q < 0$, there is no such equivalence radius.

We note that within the virial radius, the theoretical expectation embodied in Eq.~(\ref{eq:vesc}) works to high precision up to $z \sim z_{t} + \delta z$, where $z_t$ is the transition redshift and $\delta z$ is small. See, for example, Fig.~4 in Ref. \cite{miller2016}. Beyond the transition redshift, however, the analytic theory of Eq.~(\ref{eq:vesc}) is complicated both by cluster assembly dynamics and the theoretical ambiguity of what occurs to the escape velocity profile in a universe that is approaching the Einstein-de Sitter case. When $q > 0$, the internal dynamics of a bound system like a galaxy cluster are solely governed by the Poisson equation with gravity acting to source to accelerate the member galaxies.

Given this, in the Fisher matrix analysis that follows, we pick $z=0.8$ as the maximum redshift out to which we can realistically push our probe. Later on in the paper we discuss the implications of defining a redshift range for the probe.

Having described both the cluster and cosmological parameter dependence of our probe, and having considered the regimes of applicability of our probe, we now quantitatively characterize, through the Fisher matrix formalism, how well our observable may be able to constrain cosmological parameters.

%%%%%%%%%%%%%%%%
%%%%%%%%%%%%%%%%
%%% Fisher   %%%
%%%%%%%%%%%%%%%%
%%%%%%%%%%%%%%%%
\section{Fisher matrix}
The Fisher matrix formalism has been vital in predicting how well a given cosmological probe can constrain any given set of cosmological parameters \cite{1997tegmark,heavens16,coe,huterer2001,DEtaskforce}.

In this section we briefly go over the Fisher matrix formalism. We construct the Fisher matrix for our observable quantity, include parameters describing systematic errors in these observations, discuss how we apply priors on cluster and cosmological parameters, and calculate the marginalized errors on cosmological parameters.

%% formalism%%%

\subsection{Formalism}
In general, the Fisher information matrix is a function of the derivatives of the log likelihood of our observable with respect to the observable's parameters ($p$),
\begin{equation} 
F_{ij} = \bigg\langle -  \frac{\partial^2 \ln \mathcal{L}}{\partial p_i \partial p_j} \bigg \rangle.
 \label{eq:fisher_gen}
\end{equation}
Given that the observable quantity is the escape velocity as a function of redshift and radius we have that the Fisher matrix elements are sums over clusters (index $n$) and radii (indices $k$ and $l$),
\begin{equation} 
F_{ij}=   \sum_{nkl}     \frac{\partial v_{esc}(z_n,r_k)  }{\partial p_i}  (C^{-1})_{kl} \frac{\partial  v_{esc}(z_n,r_l)}{\partial p_j},  
 \label{eq:fisher}
\end{equation}
where $C$ is the covariance of escape velocity measurements at different radii. Here we assume that the escape velocity measurements in different clusters are uncorrelated, and that the measurement covariance for different radii in a cluster (matrix $C$) is independent of redshift.

After adding priors to our cluster parameters the Fisher information matrix is then given by,
\begin{equation} \label{eq:Fmatrix}
F = F_{ij} + F_{prior}.
\end{equation}
From this $F$ we can then compute the marginalized lower bound of the uncertainty on any of our $N$ parameters via the Cram\'{e}r-Rao inequality, $\sigma_{p_i} \geq \sqrt{(F^{-1})_{ii}}$. The inverse is calculated via Gaussian elimination. We ensure that the inversion of the matrices we work with are stable by calculating the condition number for each $F$ matrix and ensuring that it is less than $\sim 10^{12}.$

As detailed in the previous section, the parameters in our observable can be split up into two sets: cluster and cosmological parameters. Consequently, the parameters that we will be taking the derivatives with respect to ($p$) can be represented by the union of the following two sets:

\begin{equation}
 p \in p_{clus} \cup p_{cosmo}.
\end{equation}

The parameters describing the clusters 1 to $N$ (i.e. three parameters for the mass profile plus one more describing the anisotropy parameter, per cluster) can then be encapsulated in the following set,

\begin{equation}
 p_{clus} \in  \{ \beta_1, \alpha_1, r_{-2,1}, \rho_{-2,1}, \dots, \beta_N, \alpha_N,r_{-2,N}, \rho_{-2,N} \}.
\end{equation}

The cosmological parameter set  $p_{cosmo}$ is composed of the cosmological parameters. We  detail the cosmological models we study, and the corresponding sets of $p_{cosmo}$, in the next section.

 \begin{figure*}
\centering
  \includegraphics[width=1\textwidth]{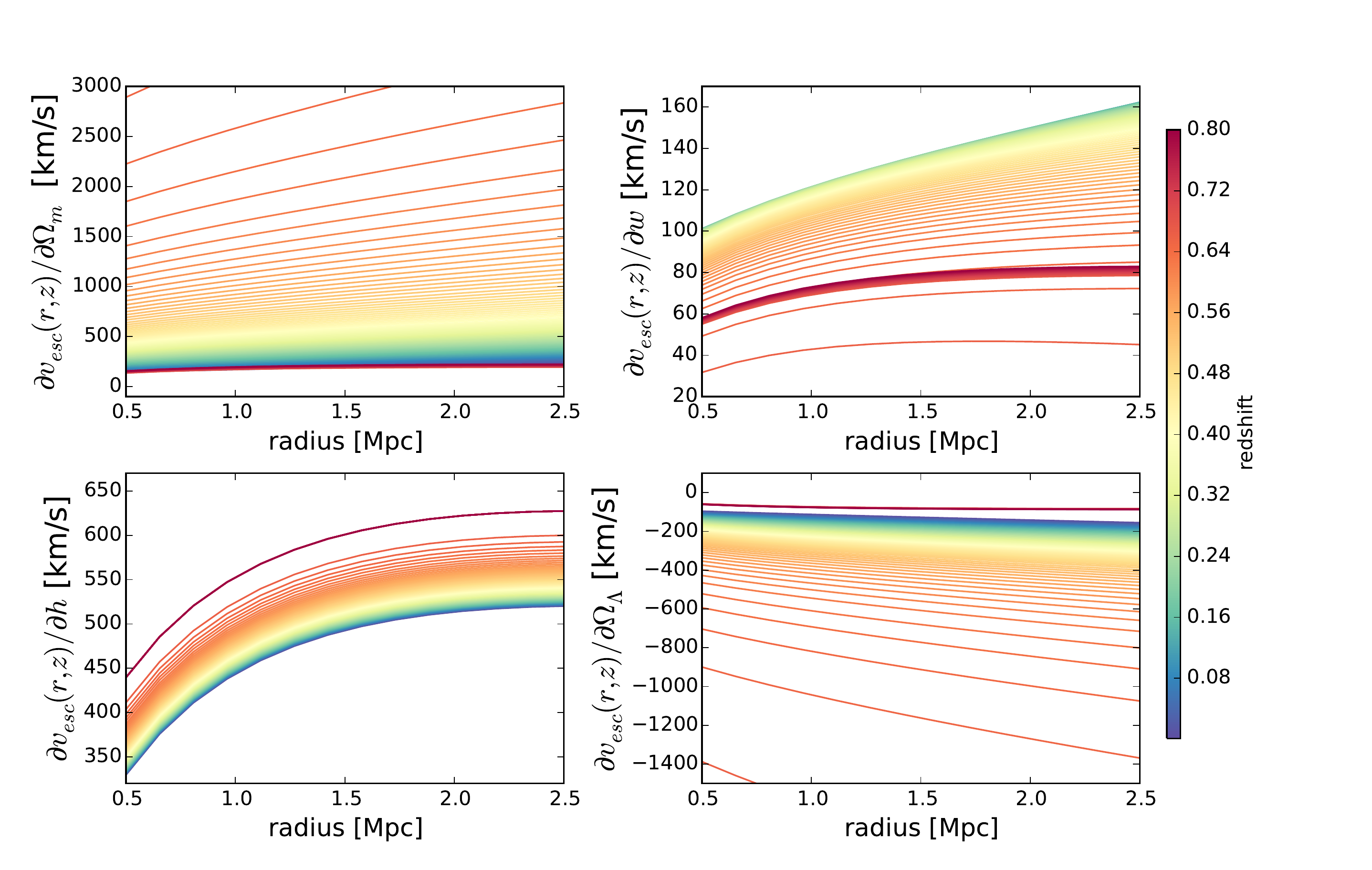}
  \caption{Sensitivity of the observable $v_{esc}(r,z)$ to cosmology.  The partial derivatives of the escape velocity profile are calculated numerically at 100 different redshifts ($z$) with respect to the various cosmological parameters for our fiducial cluster. The specific redshift of a given profile can be identified through the color bar on the right-hand side of the figure. In all cases, the information grows (i.e. the observable becomes more sensitive) the farther out we go radially. Some parameters are most sensitive at higher redshifts ($\Omega_M$, $\Omega_{\Lambda}$ and $h$) while others are more sensitive at lower redshifts ($w$). Note that beyond the transition redshift for our fiducial cosmology, the derivatives with respect to all parameters reach a limit, as implied by Section II B 2.}
\label{fig:deriv_four}
\end{figure*}

Considering these sets, we therefore have a $N_{dim}$ by $N_{dim}$ dimensional Fisher matrix $F$ given by:
\begin{equation}
 N_{dim} = 4 \times N_{clus} + N_{cosmo}
\label{Ndim}
\end{equation}
where $N_{clus}$ is the number of clusters and $N_{cosmo}$ is the number of cosmological parameters.   We provide a sketch of the Fisher matrix structure in Fig.~\ref{Fij} and explore its structure more thoroughly in the Appendix. 

\subsection{$F_{ij}$ matrix}
In this subsection we focus our attention on the components that make up the $F_{ij}$ matrix (Eq.~(\ref{eq:fisher})). 

\subsubsection{Fiducial cluster}

As with any Fisher matrix analysis, the derivatives of Eq.~(\ref{eq:fisher}) are calculated at a given set of fiducial values. For our fiducial cosmology we pick $p_{cosmo,fid} \in \{  \Omega_M= 0.3, \Omega_{\Lambda} = 0.7, w_0 = -1, w_a = 0, h= 0.7 \}$. For our fiducial cluster parameters we pick $p_{clus,fid} \in \{  \alpha = 0.1984,  \rho_{-2} = 1.0521 \times 10^{14} [M_{\odot}/\text{Mpc}^3] ,  r_{-2} = 0.497 \text{[Mpc]}, \beta = 0.145 \} $. The $\beta$ fiducial value is around what has been estimated for galaxy clusters (see Refs.  \cite{lokasabell,benatov,lemze2009,wojtak,macs1206}) and the three Einasto fiducial cluster parameters are equivalent to a cluster of mass $M_{200}= 4\times10^{14}.$  More specifically, we use the mass-concentration relation  in Ref. \cite{sereno} to map our fiducial mass $M_{200}$ to the Einasto parameters at $z=0$ by fitting the Navarro-Frenk-White (NFW) density profile to the Einasto density profile. For analytical representations of these density profiles see Ref. \cite{sereno2} and references therein. 

Furthermore, to calculate the derivatives of Eq.~(\ref{eq:fisher}) we place our fiducial cluster along different redshifts $z_n$ and recalculate its angular size $\theta $ with fixed $r$ at that given redshift via Eq.~(\ref{eq:Da}). We discuss this radial range in the next sub-subsection.

In Fig.~\ref{fig:deriv_four} we plot some of the derivatives for our fiducial cluster used in our Fisher matrix analysis. More specifically, we plot the radial derivatives of $v_{esc}(r,z)$ with respect to various cosmological parameters for 100 clusters of the same fiducial mass uniformly distributed in the range $0 \leq z \leq 0.8.$  Fig.~\ref{fig:deriv_four} is a useful way to study the sensitivity of our observable to various cosmological parameters. In particular Fig.~\ref{fig:deriv_four}  tells us that our observable is more sensitive to certain cosmological parameters at higher redshifts, such as $\Omega_M$, (the red lines, which represent high redshift clusters are higher than the blue ones, which represent low redshift clusters), and more sensitive to other parameters at lower redshift, such as the dark energy equation of state parameter $w$. Judging from the derivatives in Fig.~\ref{fig:deriv_four}, our probe is most sensitive to the parameter $\Omega_M$. Note also that for all parameters, the sensitivity increases with radius. We emphasize that beyond the transition-to-acceleration redshift ($z_t = 0.671$) in the standard $\Lambda$CDM model we no longer gain much cosmological information as encapsulated by Eq.~(\ref{eq:v_esc_limit}). Beyond this redshift we still get information cosmological information via $r$ which is a function of the angular diameter distance (Eq.~(\ref{eq:Da})).

\subsubsection{Radial bins and covariance matrix $C^{-1}$}

Having described the derivatives of Eq.~(\ref{eq:fisher}) we now 
describe the covariance matrix $C$ in that same equation. Simply put, this matrix embodies the covariance of between different measurements of our observable at a given radial bin in a cluster $n$.

For a perfect three-dimensional observation of the galaxy velocities, there would be no projection effects and therefore one would expect a nonzero covariance between $v_{esc}$ at different radii. This effect is random between different radial bins, and it effectively decouples the measurements in different radial bins, reducing the bins' covariance. More specifically, this scatter arises from the fact that when observing a galaxy cluster, random galaxies along the line of sight may be included in the phase space. This drastically reduces the covariance between radial bins.   Therefore, it is a good assumption that the radial covariance matrix is diagonal. That is, $C^{-1}$ is reduced to $\sigma_{v_{esc}}^{-2}$. 

Mathematically, at any given radius, the uncertainty in $v_{esc}(r,z)$ is given by the combination of the spectroscopic uncertainty, uncertainty due to the edge measurement, and any additional intrinsic uncertainties,

\begin{equation}
\sigma_{v_{esc}} = \sqrt{ \sigma_{spec}^2 + \sigma_{edge}^2 + \sigma_{edge,int}^2}.
\label{sigma_edge}
\end{equation}

We choose $\sigma_{spec} = 50 $ kms$^{-1}$ to match the redshift accuracy of modern spectroscopic surveys \citep{Bolton12}. To calculate the uncertainty on the statistically inferred edge at a given radial bin we follow Ref. \cite{gifford2013a} who  used simulations to show that when viewing along a line-of-sight the edge can be recovered to high statistical precision ($\sim$ 5\% or less) when 100 or more galaxies are used in the phase spaces (see their Figure 4 -- bottom right). For our fiducial cluster this corresponds to 50 kms$^{-1}$ since the observed edge is typically around 1000 kms$^{-1}$. We also allow for an intrinsic scatter between the observed edge and how accurately it can recover the true underlying gravitational potential. Ref. \cite{gifford2013b} used simulations to quantify the statistical accuracy and precision of the cluster mass using the projected edge when the density, potential, and anisotropy are known exactly. They find that there is a statistical floor of 25\% error in mass, which translates into a 12.5\% error in the edge since mass scales as the square of the potential. For our fiducial cluster, this intrinsic error corresponds to $\sim 125$ kms$^{-1}$. 

As in Eq.~(\ref{sigma_edge}), we sum these three components in quadrature giving $\sigma_{v_{esc}} = 143.61$ kms$^{-1}$ for our fiducial cluster. This is about a 15\% total error in the measurement of the projected phase space escape velocity edge. Lastly, note that if  we double the total error budget on $v_{esc}$, our constraints on the cosmological parameters increase by about 80\%.

We next tackle the question of how many radial bins we should use for a given cluster in our Fisher matrix calculations. 

Our one requirement here is to be able to resolve the shape of the velocity edge vs.\ radius, $v_{esc}(r)$. In all cases, we assume densely sampled phase spaces (i.e., 100-200 galaxies within the virial radius). Since more massive clusters are also larger in size, we can maximize the number of useful radial bins by choosing the most massive clusters for our analysis. 

At the moment, we can typically rely on reasonably accurate weak lensing mass estimates for the highest mass systems in the universe. Therefore, for this paper we will assume an SPT (South Pole Telescope)-like sample with $M_{200} > 3\times10^{14}M_{\odot}$ \cite{high2010}. More specifically, as mentioned elsewhere in the paper, we pick a fiducial mass of $M_{200} = 4 \times 10^{14} M_{\odot}$.

Furthermore, as detailed in Section IIA, we can only work within the radial range of about 0.3 - 1 virial radii if we assume that the anisotropy profile can be reduced to a single parameter. Henceforth for our profiles and constraints we pick a radial range between $0.5 \leq r \leq 2.5$ Mpc, the outer range corresponding to about $r_{200}$ (where the density reaches 200 $\times$ the mean value) and the inner range corresponding to about  $0.3 \times r_{200}$ described above \cite{high2012}. This radial range, given $\Delta r \sim 0.1$ Mpc yields $N_{bins}=14$ radial bins. This is what we use in our calculations throughout our paper (see Eq.~(\ref{eq:trueFmatrix}) below).  

Note that if we change the number of radial bins from 14 to 7 and 21, the constraints on the cosmological parameters change by $\sim 20\%$. We can therefore in principle get better constraints than what is presented in the next section by increasing the number of radial bins. However, binning too finely is not desirable given that it can introduce additional statistical noise in the observable and may not even be possible, given the density of galaxies in the phase spaces that are observationally viable.

Given all of this, Eq.~(\ref{eq:fisher}) thereby becomes,

\begin{equation} 
F_{ij}=  \sum_{n=1}^{N_{clus}} \sum_{k = 1}^{N_{bins}}  \frac{1}{\sigma_{v_{esc}}^2}     \frac{\partial v_{esc}(r_k,z_n)  }{\partial p_i}   \frac{\partial  v_{esc}(r_k,z_n)}{\partial p_j}.
\label{eq:trueFmatrix}
\end{equation}
and is therefore a sum over $N_{clus}$ clusters and $N_{bins}$ radial bins. This is the $F_{ij}$ matrix we utilize for all of our constraints presented in section IV, in conjunction with the prior matrix which is discussed in the following subsection.

%% priors%%%
\begin{table*}[]
\centering
\resizebox{1\textwidth}{!}{%
\begin{tabular}{@{}lccccl@{}}
\toprule
\\ \hline \hline 
\multicolumn{6}{c}{\textbf{Cluster Parameter Uncertainties}}                                                                                                                                                                \\ \hline \hline \midrule
\textbf{WL mass error}              & \textbf{$\sigma_{\alpha}$} & \textbf{$\sigma_{\rho_{-2}} [M_{\odot}/\text{Mpc}^3]$} & \textbf{$\sigma_{r_{-2}}${[}Mpc{]}} & \textbf{$\sigma_{\beta}$} & \textbf{$\sigma_{v_{esc}}$ {[}kms$^{-1}${]}}  \\ 
{5\% stat +  5\% cosmo sys ("stacked") }   & 0.0024                     & $5.887 \times10^{12}$                                  &  0.0314                               & 0.02                      & 90.14                            \\
{20\% stat +  20\% cosmo sys ("40\% mass scatter")}  & 0.0096                     & $23.589 \times10^{12}$                                 &  0.1342                              & 0.5                       & 143.61                           \\
{40\% stat +  40\% cosmo sys ("80\% mass scatter")}  & 0.0181                    & $43.904 \times10^{12}$                                 & 0.2913                               & 0.5                       & 143.61                           \\
{40\% stat  ("Riess et al 2016 prior on $h$")} & 0.0096                       &   $23.589 \times10^{12}$                                 & 0.1342                               & 0.5                       & 143.61                           \\ \cmidrule(r){1-6}
\hline \end{tabular}%
}
\caption{Cluster parameter uncertainties that make up the $F_{prior}$ matrix used in the various cases considered in the Constraint Forecasts section as well as the error on the edge $\sigma_{v_{esc}}$ that makes up the $F_{ij}$ matrix. Note that in principle there is no covariance between $\beta$ and the three other parameters, but we simply change the uncertainty on $\beta$ for the other cases with reduced weak lensing mass scatter. Furthermore, as explained in the text, we use the weak lensing (WL) mass percent error on $M_{200}$ as shorthand to describe uncertainties in all three Einasto parameters. In the last row we tabulate the uncertainties in the three Einasto parameters after applying a prior on $h$ from from Ref. \cite{riess2016}. All other cluster parameter uncertainties listed contain both statistical error as inferred from weak lensing (WL) analyses and systematic error from cosmology as explained in the "Prior information" section and Appendix C.}
\label{table1}
\end{table*}

\subsection{Prior information}
The Fisher information matrix formalism allows us to add additional independently measured information attained on certain model parameters (both cluster and cosmological) to our Fisher matrix. This is implemented via the prior information matrix $F_{prior}$ in Eq.~(\ref{eq:Fmatrix}). The structure of this matrix is given by,

\begin{equation}
F_{prior} = \left(
 \begin{array}{ccccc}
  C_{cosmo}^{-1} & &  &   \text{\huge0} \\
    & C_{cluster}^{-1} &  & \\
    & & &  C_{cluster}^{-1} &  \\
      \text{\huge0}   &  & &  & \ddots
 \end{array}\right).
 \label{eq:Fpriorstructure}
\end{equation}

We discuss the elements of this prior information matrix below.

\subsubsection{Cluster prior information ($C_{cluster}^{-1}$)}
In our case, the mass parameters (which come from weak lensing mass estimates) and the anisotropy parameter (which comes from analysis of the phase spaces of the clusters via the Jeans equations) are known to within some precision from these independent measurements. We therefore add a prior information matrix to account for this external information on the non-cosmological parameters.

More specifically, the error bars on the weak lensing mass estimates are chosen to be similar to what is reported in the literature based on recent observations.  As a representative sample, see the metacatalog compiled in Ref. \cite{sereno}. We particularly choose an $M_{200}$ error range from  20\% to 40\% which is based on ground-based imaging. For instance, Ref. \cite{applegate} reports typical statistical errors of $\sim$ 20\% using Surprime-Cam imaging for 50 clusters to a redshift of 0.7. Similarly, Ref. \cite{umetsu} primarily used Subaru/Suprime-Cam to obtain weak lensing errors at a level of $20-30\%$ based on the CLASH sample. Ref. \cite{melchior} also reports $\sim$40\% statistical errors for four clusters using science verification data from the Dark Energy Survey on the CTIO 4m Blanco DECam imager. This is why, as our baseline, we use the upper range of 40\% statistical error on the mass. In what follows we also consider 5\% statistical error as a floor that can be potentially achieved through the technique of stacking galaxy clusters \cite{becker2011,Rozo2011}.

However, there are also systematic errors that need to be considered. These can come from a variety of observational sources including the telescope point-spread function, the background redshift distribution, the intrinsic shape variations of the background galaxies, and more. Ref \cite{applegate} reports systematic errors that are small compared to the statistical errors (7\%) when the telescope optical system is well characterized. 

Few researchers have allowed for cosmology to vary during the weak lensing mass estimation process. As we approach higher precision requirements, we will need to incorporate variations in cosmology when calculating the cluster masses and it could play an important role in the overall error budget. Therefore, we estimate how large the mass errors would grow when allowing cosmology to vary during the mass estimation process. We do this while keeping the nominal statistical error, which represents the current conservative end of ground-based results. As explained in the Appendix C, we find that the total (statistical + cosmological systematic) error, for example, increases from 20\% to 40\%. That is, considering cosmological systematics in the weak lensing analysis increases the mass error by a factor of 2. Also as detailed in the Appendix, the cosmological systematics can be undercut by applying a prior on the Hubble parameter $h$. We tabulate these results in Table 1.

Note that when in this paper we speak of some percentage of "mass scatter" we refer to both the statistical and cosmological systematic error as tabulated in the first three rows of Table 1. The last row of Table 1 shows the uncertainties for a case in which the Hubble prior has been implemented. In that case, the 80\% uncertainty on the mass is reduced to 40\% after applying the prior on $h$ from Ref. \cite{riess2016} ($\sigma_h = 0.0174$). See section C in the Appendix for details.

Note also that when we speak of mass scatter  we refer to how the statistical uncertainties in all three cluster parameters change ($\alpha$, $\rho_{-2}$ and $r_{-2}$). More specifically, we use the more widely reported percent scatter on $M_{200}$ as a shorthand to describe the uncertainties in our Einasto cluster parameters. As with the fiducial mass, we attain these uncertainties by propagating the $1\sigma$ errors on the NFW density profile to the Einasto density profile parameters by fitting the profiles out to one virial radii.

Also unless otherwise stated, and as shown in Table 1, for the anisotropy parameter we use an uncertainty of  $\sigma_{\beta} = 0.5$ which is also a fairly conservative choice. For example, see uncertainties on $\beta$ as derived from a Jeans' analysis in Ref. \cite{lokasabell}. 

These parameter uncertainty priors come into our prior matrix only along the diagonal. This is because we are already considering covariances on the prior information between cluster and cosmological parameters through the aforementioned analysis. Note that this does not mean that we are not considering covariances between parameters. We are, these are all encapsulated in the Fisher matrix $F_{ij}$. We simply neglect covariance on most prior information. In this sense, our attempt to grapple with the covariances in our prior matrix (detailed in Appendix C) is only an approximation. Note, however, that some parameter uncertainties in the prior matrix are actually nil. For instance, a Jeans equation analysis inference of $\beta$ is what allows us to decouple the edge uncertainty $\sigma_{v_{esc}}$, as well as other cluster parameters from $\sigma_{\beta}.$ In contrast, inferring $\beta$ from edge measurements as done  by Ref. \cite{starkApJ} would introduce a complete covariance between edge measurements and the anisotropy parameter. The Jeans equation evades this problem given that it is a function of the derivative of the potential rather than the potential itself.

Lastly, the inverse cluster covariance matrices ($C_{cluster}^{-1}$) contain the aforementioned priors on the mass parameters as well as the prior on the cluster anisotropy parameter. We  add covariance between relevant parameters of the cluster prior submatrix. In particular, we model the covariance between cluster parameters $r_{-2}$ and $\rho_{-2}$. We discuss the structure of the submatrices $ C_{cluster}^{-1}$ in the Appendix.

\subsubsection{Cosmological prior information ($C_{cosmo}^{-1}$)}

In relation to the cosmological prior information matrices, we note that for most cases we are not adding any cosmological priors so that in Eq.~(\ref{eq:Fpriorstructure}),  $C_{cosmo}^{-1} $ is an $N_{cosmo} $ by $N_{cosmo} $ null matrix. Where noted, we add a diagonal prior on the $C_{cosmo}^{-1}$ matrix from the 2.4\%  determination of the Hubble constant from Ref. \cite{riess2016}.

%%%%%%%%%%%%%%%%
%%%%%%%%%%%%%%%%
%%% Forecasting constraints  %%%
%%%%%%%%%%%%%%%%
%%%%%%%%%%%%%%%%
\section{Constraint Forecasts}
We now derive marginalized uncertainties on  cosmological parameter $p_{cosmo}$, marginalized over the cluster nuisance parameters $p_{cluster}$. Note that in some cases we also marginalize over remaining cosmological parameters to generate 2-dimensional likelihoods. 

We consider three cosmological models. The first two assume a flat universe ($\Omega_k = 0 $) and the last case assumes a non-flat universe ($\Omega_k \neq 0$).
%%%%%%%%%%%%%%%%
%flat wCDM
%%%%%%%%%%%%%%%%

\subsection{Constant equation of state $w$}
\label{qH2}
For this case, we consider the following set of cosmological parameters,  
\begin{equation}
    p_{cosmo} \in \{ \Omega_M, w, h\}.
\end{equation} 
Via Eq.~(\ref{eq:hubbleparam}) we find the Hubble parameter which yields 
\begin{equation} 
H^2 = H_0^2 E(z)^2 \\
= H_0^2 \bigg[\Omega_M (1+z)^3 + \Omega_{DE} (1+z)^{3(1+w)}\bigg].
\end{equation}
And the deceleration parameter via Eq.~(\ref{eq:qfunction}),
\begin{equation} 
q = \frac{1}{2} \bigg[ \Omega_M(z) + (1+3w) \Omega_{DE}(z) \bigg].
\end{equation}
Here,  $\Omega_M(z) =  \Omega_M (1+z)^{3} E(z)^{-2} $ and $ \Omega_{DE}(z)= \Omega_{DE} (1+z)^{3(1+w)} E(z)^{-2}$. The combination then of the deceleration parameter and Hubble parameter that makes up the equivalent radius and our observable is given by,
\begin{equation} 
qH^2 = \frac{ H_0^2}{2} \bigg[\Omega_M (1+z)^{3} + (1+3w)(1-\Omega_M) (1+z)^{3(1+w)} \bigg].
\end{equation}
Notice that the quantity $E(z)$  cancels out in this expression. 

\begin{figure}
  \centering
  \includegraphics[width=1\linewidth]{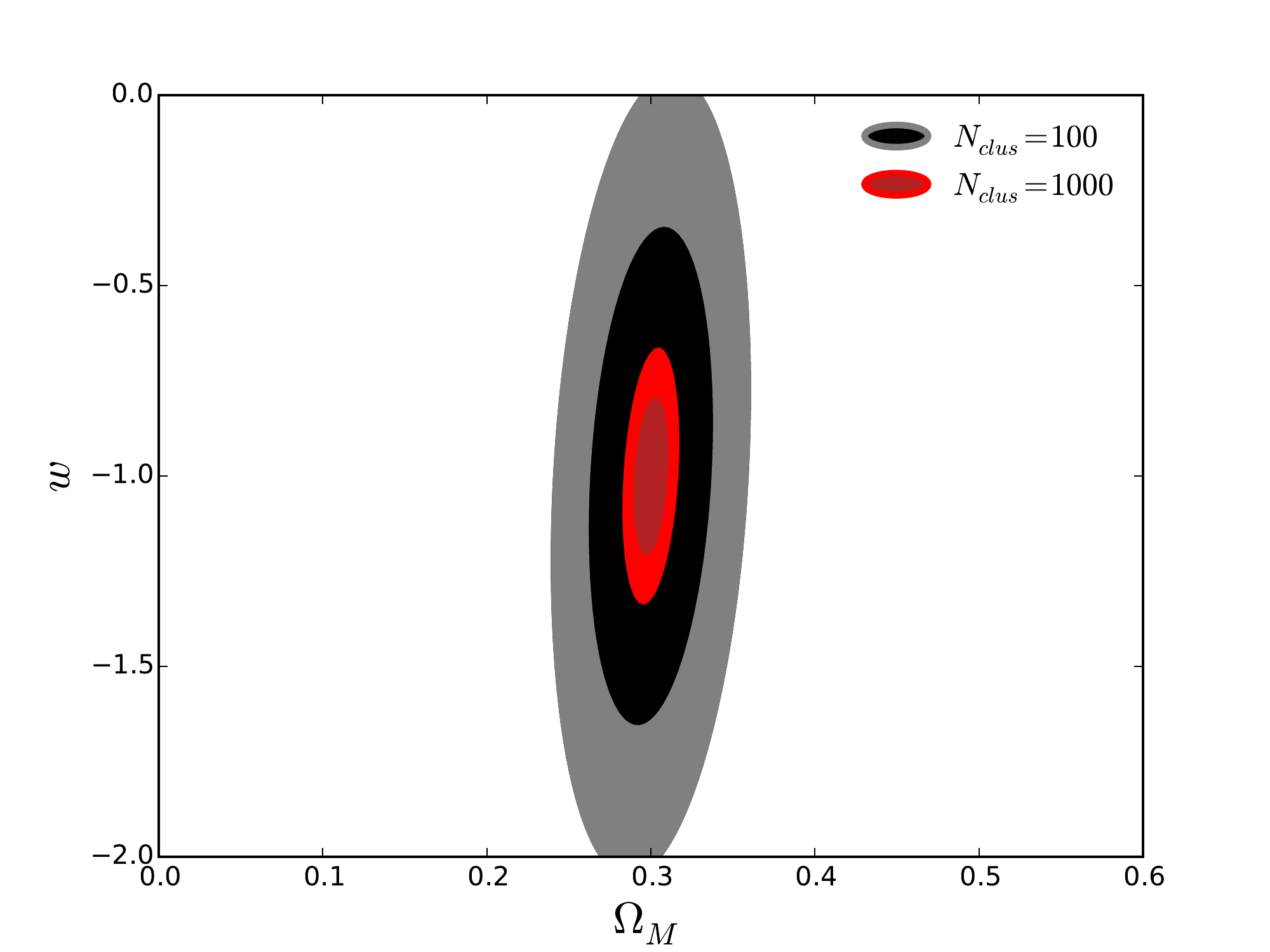}
  \caption{68\% and 95\% confidence constraints for the flat $w$CDM case after marginalizing over all other parameters. We use $N_{clus}=  1000$ (100) clusters as shown in red (black) uniformly distributed in the range $0 \leq z \leq 0.8$. The priors on the Einasto parameters assume a uniform $80\%$ weak lensing mass scatter for all redshifts (see Table 1). The  1$\sigma$ errors are  $\sigma_{\Omega_M} =  0.007 (0.025)$ and $\sigma_w = 0.138 (0.431)$ for $N_{clus}= 1000$ (100).}\label{om_w}
\end{figure}

The constraints in the $\Omega_M - w$ plane, after marginalizing over $h$ and all other cluster parameters are shown in Fig.~\ref{om_w}. In particular, in black/grey (dark/light red) we plot the 68\% and 95\% confidence level constraints for a set of $N_{clus} = 1000 (100)$ clusters uniformly distributed in the range $0 \leq z \leq 0.8$. The  marginalized 1$\sigma$ errors are  $\sigma_{\Omega_M} =  0.007 (0.025)$ and $\sigma_w = 0.138 (0.431)$ for $N_{clus}= 1000$ (100).  This is an extraordinarily tight constraint considering that it is achieved by the escape velocity method alone, before adding constraints from other probes.

%%%%%%%%%%%%%%%%
%flat w(z)
%%%%%%%%%%%%%%%%

\subsection{$w_0$ and $w_a$}
In the previous case we considered a constant dark energy of equation of state (EoS).  However, in principle the dark energy EoS can evolve with redshift. A popular way to parametrize the redshift evolving dark energy EoS through the so-called Chevallier-Polarski-Linder (CPL) parametrization given by, (see Refs. \citep{chevallierandpolarski,linder})

\begin{equation}
w(z) = w_0 + w_a \frac{z}{1+z}.
\end{equation}

In this case we consider the following set of cosmological parameters,
\begin{equation}
p_{cosmo} \in \{ \Omega_M, w_0, w_a, h\}.
\end{equation}
 Again we derive the Hubble parameter for this case,
\begin{multline} 
H^2 =  H_0^2 E(z)^2\\
=H_0^2 \Bigg[ \Omega_M (1+z)^{3}  + \Omega_{DE}(1+z)^{3(1+ w_0 + w_a) } e^{- 3w_a \frac{z}{1+z}}  \Bigg].
\end{multline} 
and the deceleration parameter
\begin{multline} 
q =  \frac{1}{2} \Bigg[ \Omega_M(z)  + \Omega_{DE}(z)  \bigg(  1+ 3w_0 + \frac{3w_a z}{1+z}  \bigg) \Bigg].
\end{multline} 
The redshift evolving mass and dark energy densities are given by, $\Omega_M(z) =  \Omega_M (1+z)^{3} E(z)^{-2} $ and $\Omega_{DE}(z)=  {\Omega_{DE}} (1+z)^{3 (1+ {w_0}+{w_a}) }e^{-\frac{3 {w_a} z}{1+z}}  E(z)^{-2} $.

Again the redshift evolving energy density term $E(z)$ cancels and we are left with, 
\begin{multline} 
qH^2 =  \frac{H_0^2}{2} \Bigg[ \Omega_M (1+z)^3  + (1-\Omega_M) (1+z)^{3 (1+ {w_0}+{w_a}) }  \\
\times \exp \bigg\{{-\frac{3 {w_a} z}{1+z}}\bigg\} \bigg(  1+ 3w_0 + \frac{3w_a z}{1+z}  \bigg)  \Bigg].
\end{multline} 

\begin{figure}
  \centering
  \includegraphics[width=1\linewidth]{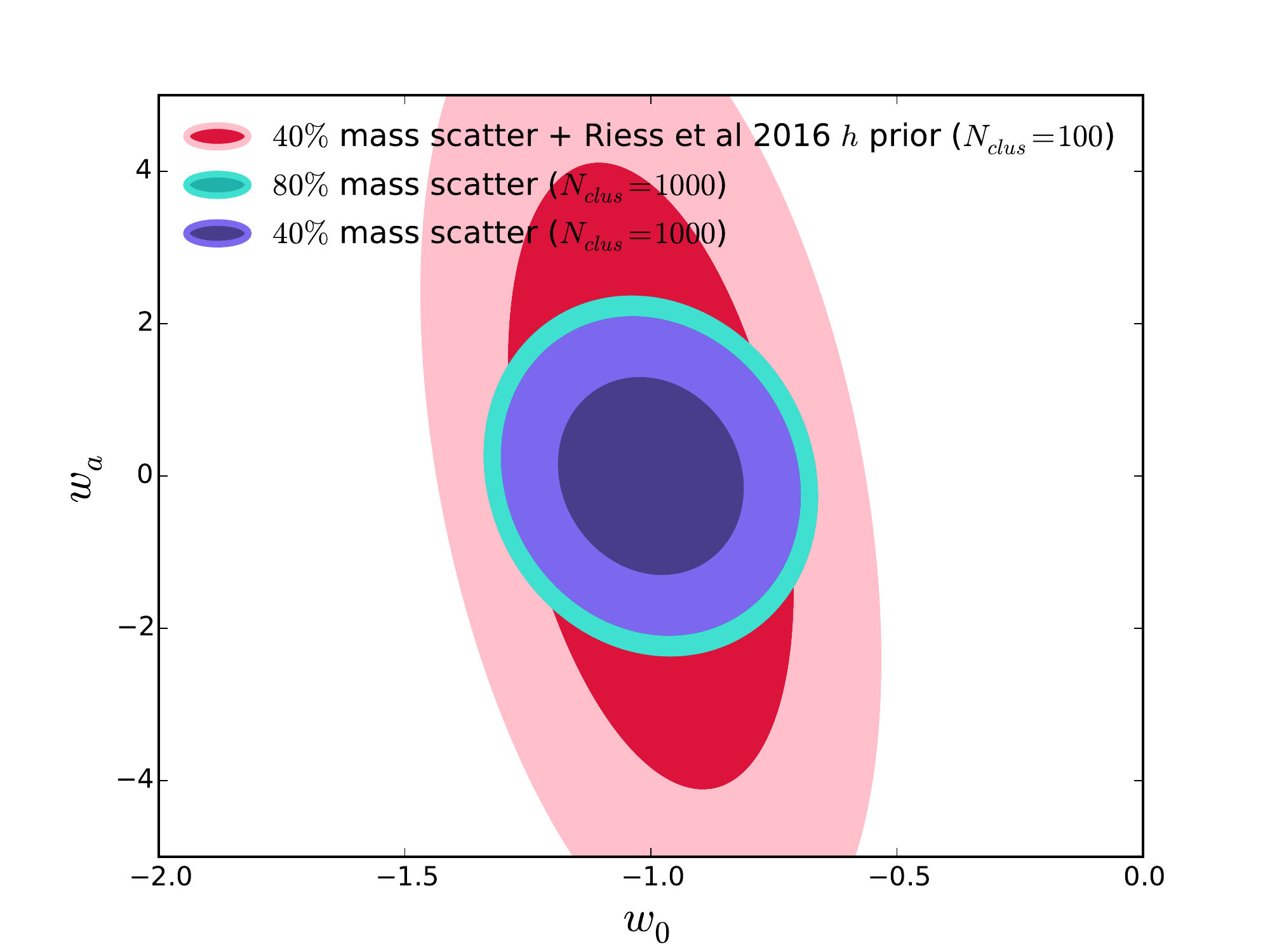}
  \caption{68\% and 95\% contours for the dynamic dark energy case using the CPL parametrization of dark energy marginalized over $\Omega_M$ and $h$ as well as the other cluster parameters. More specifically, we use $N_{clus}=  1000$  clusters uniformly distributed between $0 \leq z_c \leq 0.8$ with $80\%$ ($40\%$)  weak lensing mass scatter in the  turquoise (purple) contour which yields:    $\sigma_{w_0} = 0.139 (0.124)$  and $\sigma_{w_a}= 0.968 (0.857).$ With $N_{clus}=100$ uniformly distributed in the same redshift range as before and with $40\%$ mass scatter but now adding a prior on the Hubble constant from Ref. \cite{riess2016} we attain, $\sigma_{w_0} = 0.191 $ and $\sigma_{w_a}= 2.712 $ (pink contours).}\label{w0_wa}
\end{figure}

We show the resulting 2-dimensional $w_0 - w_a$ likelihood in Fig.~\ref{w0_wa} for a uniform set of clusters in the range $ 0 \leq z \leq 0.8$, after marginalizing over $\Omega_M$, $h$, and all other cluster parameters. With $N_{clus}=  1000$  clusters uniformly distributed between $0 \leq z_c \leq 0.8$ with $80\%$ ($40\%$)  weak lensing mass scatter the turquoise (purple) contours express the following marginalized uncertainties: $\sigma_{w_0} = 0.139 (0.124)$  and $\sigma_{w_a}= 0.968 (0.857).$ 

We then reduce the number of clusters to $N_{clus}=100$ and apply a cosmological prior on the Hubble constant from Ref. \cite{riess2016} ($\sigma_h = 0.0174 $) to yield the following marginalized $1\sigma$ errors, $\sigma_{w_0} = 0.191$ and $\sigma_{w_a}= 2.712 $  (see pink contours in Fig.~\ref{w0_wa}). Note that the pink contour is made using the same redshift range ($0 \leq z \leq 0.8$) and systematic error  (40\% weak lensing mass scatter and $\sigma_{\beta} = 0.5$) as before but with only 100 clusters. Note that this constraint on $w_0$ is comparable to the constraint achieved with 1000 clusters of  Fig.~\ref{om_w} (red contours).

%%%%%%%%%%%%%%%%
% non flat LCDM
%%%%%%%%%%%%%%%%

\subsection{non-flat universe,  $\Omega_{M}$ and $\Omega_{\Lambda}$ }
So far we have only considered flat universes in our analysis. We now drop this assumption and assume the possibility of nonzero curvature, while fixing the dark energy equation of state to $w = -1$. We then have the following set of parameters to constrain,  

\begin{equation}
p_{cosmo} \in \{ \Omega_M, \Omega_{\Lambda}, h \}
\end{equation}

For this case the Hubble parameter is given by
\begin{equation} 
H^2 =  H_0^2 E(z)^2 = H_0^2 \Bigg[ \Omega_M (1+z)^{3}  + \Omega_{\Lambda} + \Omega_k (1+z)^2  \Bigg].
\end{equation} 
And the deceleration parameter,
\begin{equation} 
q =  \frac{1}{2} \Omega_M(z) - \Omega_{\Lambda}(z).
\end{equation} 
Multiplying these two we have,
\begin{equation} 
qH^2 =   \bigg[\frac{1}{2} \Omega_M(1+z)^3 - \Omega_{\Lambda}\bigg] H_0^2. 
\end{equation} 

\begin{figure}
  \centering
  \includegraphics[width=1\linewidth]{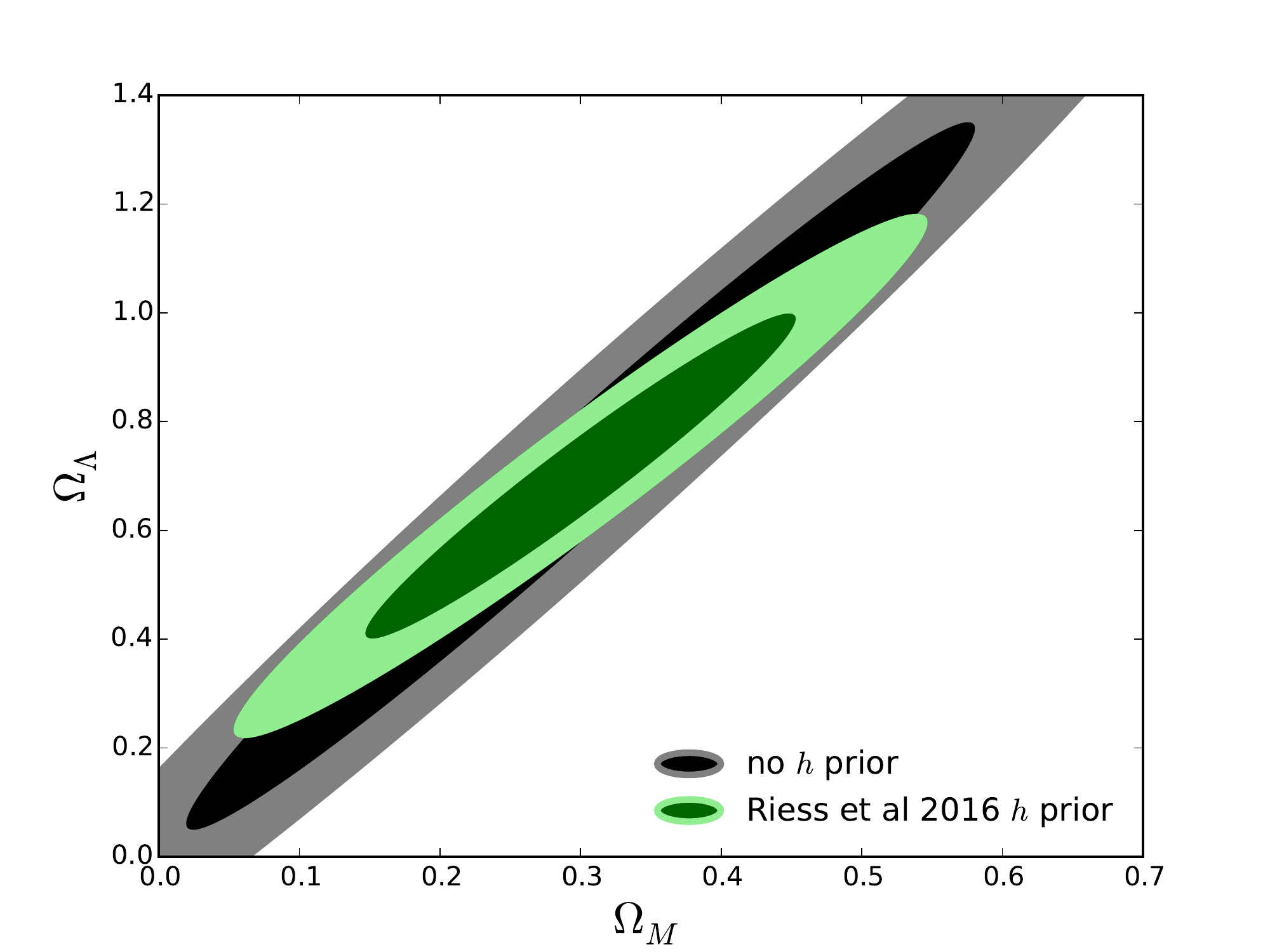}
  \caption{68\% and 95\% contours for the non-flat $\Lambda$CDM case for $N_{clus}=  100$ uniformly distributed between $0 \leq z \leq 0.8$ with 80\% mass error. Applying a prior on the Hubble constant $h$ from Ref. \cite{riess2016} ($\sigma_h  = 0.0174$) allows us to break the degeneracy and thereby significantly improve our constraints.  The marginalized $1\sigma$ constraints derived from the green (black) contours are $\sigma_{\Omega_M} =  0.101 (0.185)$and $\sigma_{\Omega_{\Lambda}} =   0.197 (0.428)$. }\label{oM_oL}
\end{figure}

The constraints in the $\Omega_{M} - \Omega_{\Lambda}$ plane,  after marginalizing over all cluster parameters are shown in Fig.~\ref{oM_oL}. With just $N_{clus} = 100$ uniformly distributed between $ 0 \leq z \leq  0.8 $ we can achieve the following marginalized uncertainties  $\sigma_{\Omega_M} =  0.101 (0.185)$ and $\sigma_{\Omega_{\Lambda}} =   0.197 (0.428)$ after applying the $2.4 \%$ level prior on $H_0$ from Ref. \cite{riess2016} (compare green to black contours). 

%%%%%%%%%%%%%%%%
%%%%%%%%%%%%%%%%
%%% obs. strategy  %%%
%%%%%%%%%%%%%%%%
%%%%%%%%%%%%%%%%
\section{Observational strategies}\label{sec:obs_strategies}
In order for the escape velocity measurements to yield competitive cosmological constraints in both the near term and the far future, several considerations must be taken into account. In this section we particularly focus on how future surveys should target specific redshifts in order to optimize cosmological constraints. We also explore the extent to which reducing systematic uncertainties in both weak lensing mass estimates and measurements of the anisotropy parameter will yield significantly better cosmological constraints when compared to simply increasing $N_{clus}$.

%%%%%%%
%%%%%%%
\subsection{Redshift range}
%%%%%%%
%%%%%%%

 \begin{figure*}
\centering
  \includegraphics[width=1\textwidth]{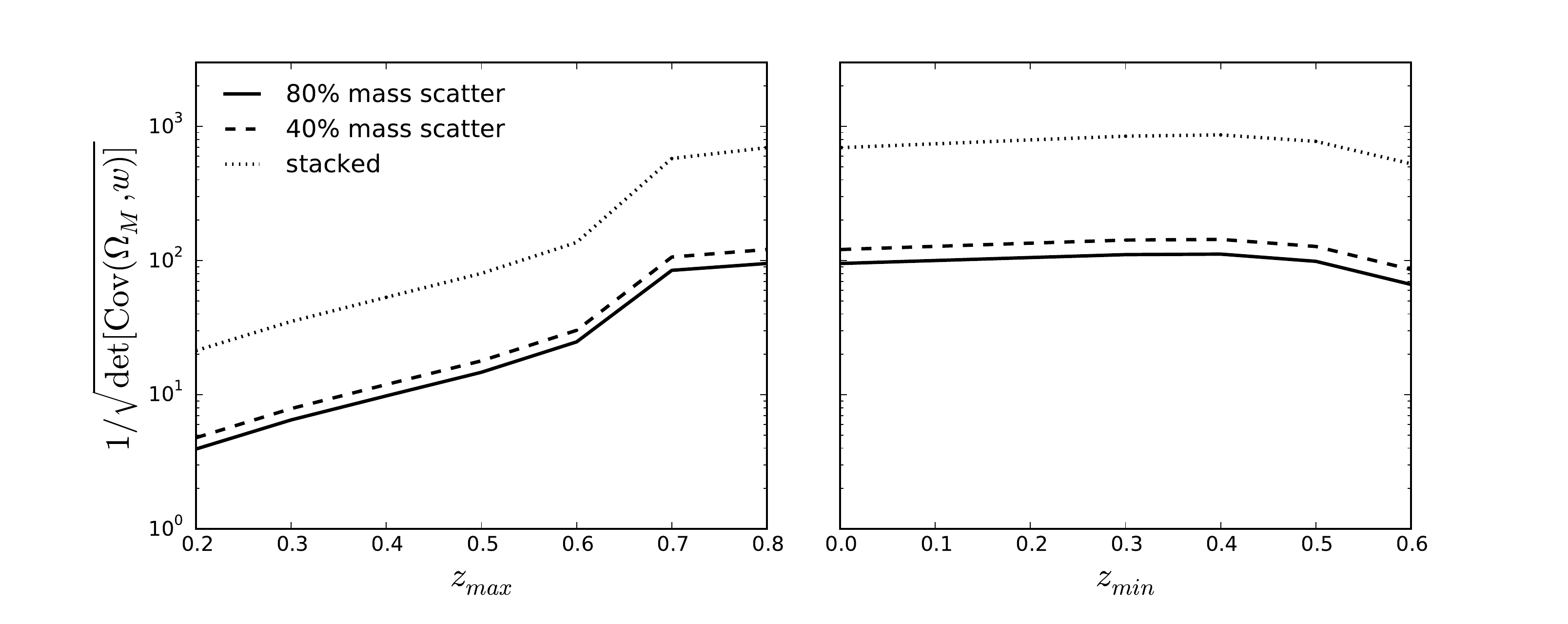}
  \caption{Inverse area of the $\Omega_M$ - $w$  covariance matrix after marginalizing over all other parameters as a function of maximum (left panel) and minimum (right panel) redshift used. We assume a uniform redshift distribution of clusters in the range $0 \leq z \leq z_{max}$ (left panel) and  $ z_{min} \leq  z \leq 0.8$ (right panel). Each calculation of the inverse area assumes a fixed number of clusters, $N_{clus} = 100$. Note that  $\Omega_M - w$ constraints can be improved by reducing the mass uncertainty from 80$\%$ (solid line) to 40$\%$ (dashed line), in which case the contour area decreases by a factor of $\sim$1.3. The constraint can be further increased by utilizing stacked weak lensing mass estimates and stacked phase spaces, and this yields a 10\% mass scatter and $\sigma_{\beta}=0.02$ (see Table 1), decreasing the contour area by a factor of $\sim 7.5$ (compare solid to dotted lines). The left panel also illustrates that using clusters beyond the transition redshift leads to a gradual loss of cosmological information. While a tighter constraint can be achieved by incorporating higher redshift clusters, the right panel demonstrates that we still need low redshift clusters to achieve the tightest constraints on $w$. See Fig.~\ref{fig:sigma_zmin}. We conclude that as broad as possible redshift range of clusters be used (e.g.\ $ 0 \leq z \lesssim 0.8$).}
\label{fig:inv_area}
\end{figure*}

\begin{figure}
  \centering
  \includegraphics[width=1\linewidth]{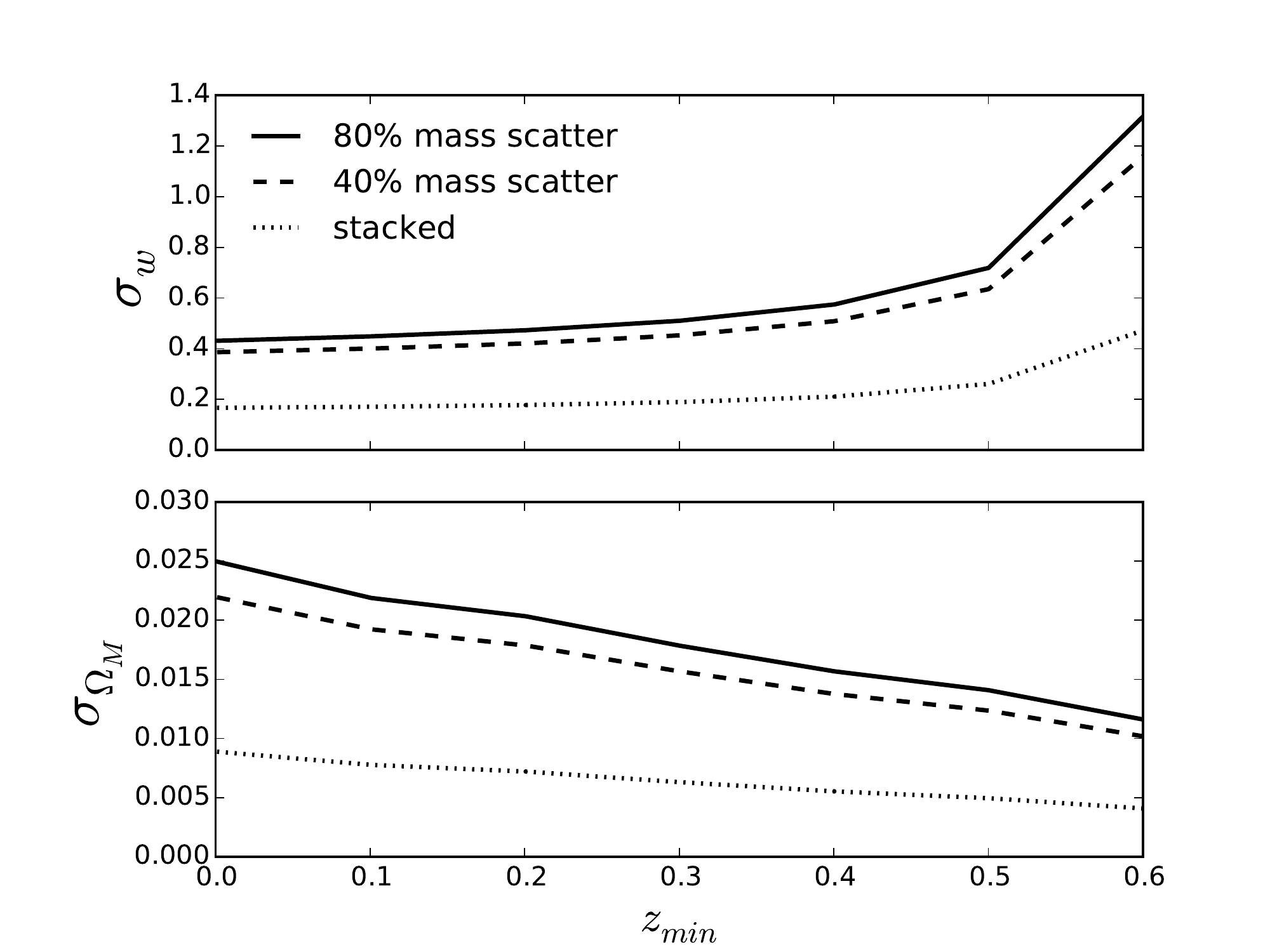}
  \caption{Marginalized $\Omega_M$ and $w$ uncertainties for the flat $w$CDM case with 100 clusters distributed in the range $ z_{min} \leq z \leq 0.8$. Note that as the minimum redshift $z_{min}$ increases a factor of $\sim$1.7 improvement in $\sigma_{\Omega_M}$ is attained (see solid  line, bottom panel). The trade-off is a significant loss of information on the dark energy equation of state parameter $w$ at high redshift (top panel). These effects combine to make the inverse area plot as a function $z_{min}$ relatively flat (see right-hand side panel in Fig.~\ref{fig:inv_area}). Note also that a maximum factor of $\sim 4 $ improvement in the uncertainty of these parameters may be achieved if cluster parameter uncertainties are reduced (compare solid to dotted lines).}\label{fig:sigma_zmin}
\end{figure}

We want to investigate how the cosmological constraints vary with a change of the limits on this redshift distribution. In particular, note that the contours shown in Figures 5-7 utilize a uniform redshift distribution in the range $0 \le z \le 0.8$  for the cluster sample. How would these contours change if we changed the cluster redshift range used? 

To quantify the effect, we calculate how the area of the $1\sigma$ contours in any given 2-dimensional cosmological parameter space changes as a function of the redshift distribution chosen. Mathematically, this entails taking the covariance matrix that contains the marginalized parameters we are interested in and calculate how its determinant changes as a function of the maximum and minimum redshifts for a given distribution.  More explicitly, the inverse area corresponding to the 2 by 2 covariance matrix for the marginalized parameters within the $1\sigma$ bound is given by (see Refs. \cite{huterer2001,DEtaskforce}), 

\begin{equation}
\text{A}^{-1}(p_i, p_j) = \frac{1}{\sqrt{|\text{det}[\text{Cov}(p_i, p_j)]|}}.
 \label{eq:invArea}
\end{equation}
For the $w_0 - w_a$ case the inverse area is the Dark Energy Task Force "Figure of Merit" (FoM) \cite{DEtaskforce}.

The result for the flat $w$CDM case as a function of redshift range used is shown in Fig.~\ref{fig:inv_area}. It shows the inverse area for a fixed number of clusters (100 in this case) uniformly distributed in the range $0 \leq z \leq z_{max}$ (left panel) and  $ z_{min} \leq  z \leq 0.8$ (right panel). 

As the left panel of Fig.~\ref{fig:inv_area} implies, we can get the tightest constraints on this cosmological parameter subset by picking 100 clusters uniformly distributed in the range $ 0 \leq z \leq 0.8$. The physical reason for this can already be inferred from Fig.~\ref{fig:deriv_four} which shows the derivatives of our observable with respect to the various cosmological parameters. In that figure we note that while our observable is sensitive to $\Omega_M$ at high redshifts, it is simultaneously more sensitive to $w$ at low redshifts. Moreover, our probe is relatively more sensitive to $\Omega_M$ than to $w$ (compare the absolute maximum of the derivatives). This immediately implies that the higher redshift clusters will end up contributing more to the joint constraint. 

However, we want to emphasize that this does not mean that we should therefore only pick high redshift clusters. In Fig.~\ref{fig:sigma_zmin} we plot the marginalized uncertainty for both $w$ (top panel) and  $\Omega_M$ (bottom panel) as a function of $ z_{min}$. As we pick higher redshift clusters the constraint on $\Omega_M$ improves but the constraint on $w$ is degraded. This is also shown by the relatively flat but ultimately decreasing  tendency of the right panel in Fig.~\ref{fig:inv_area}. As such, we emphasize that we need both high and low redshift clusters if we are going to attain a tight constraint on both $\Omega_M$ and $w$. This applies to other constraints as well.

Secondly, note that this particular optimized choice (ie. picking clusters uniformly distributed in the range $0 \leq z \lesssim 0.8$) arises from our fiducial cosmology which yields $z_t = 0.671.$ Ref. \cite{riess2007} for instance, finds a transition redshift of $z_t= 0.43 \pm0.07$ based on a linear parametrization of $q(z).$ Therefore, the inferences of the transition redshift are highly model dependent \cite{shapiro}. What this means is that, observationally, since the optimization of our probe is loosely based on the transition redshift, we recommend that a redshift distribution as broad as possible be used. In particular, we recommend that clusters uniformly distributed in the range $0 \leq z \leq 0.8$ be used. Picking this redshift range allows us to safely take into account current uncertainties in the transition redshift. 

Lastly, note also that this upper limit ($z = 0.8$) is also set by the particular processes of cluster assembly. In other words, beyond this redshift our analytic model is unable to take into account the full complexity of cluster-formation dynamics because clusters are still assembling at that redshift for acceptable cosmologies (see Fig. 2 in Ref. \cite{holz}). 

%%%%%%%
%%%%%%%
\subsection{Reducing Systematics and Stacked Clusters}
%%%%%%%
%%%%%%%
We now study the effects of reducing statistical errors on the cluster parameters. Clearly, reducing statistical errors in the weak lensing mass estimates, in the inference of anisotropy parameter, and in the measurement of the edge, will improve our cosmological constraints, but by how much? For this exercise, we consider increased precision from better measurements on individual clusters, increased cluster sample sizes, and through stacking techniques. We note that stacking is not necessarily equivalent to averaging over a large sample. For additional information on stacking, we refer the reader to detailed analyses of stacking weak lensing data and stacking phase-spaces \citep{Rozo2011, Gifford2017}. To quantify the improvements, we use  Eq.~(\ref{eq:invArea}) with a fiducial sample of $N_{clus} = 100$ and focus on the $\Omega_M - w$ case. 

Fig.~\ref{fig:inv_area} shows how constraints may be improved by decreasing the scatter on the mass parameters from 80\% (solid black line) to 40\% (dashed black line). The difference in the inverse area size is a relatively modest factor of $\sim 1.3.$ However, simultaneously reducing the uncertainty of the mass parameters to 10\% as well as reducing the uncertainty of the anisotropy parameter $\sigma_{\beta}$ and the uncertainty on the escape velocity edge $\sigma_{v{esc}}$ yields an area that is $\sim 7.5$ times smaller (compare solid lines to dotted lines).  For the exact values of the uncertainties used in our matrix for this "stacked" case see Table 1. Fig.~\ref{fig:sigma_zmin} follows and demonstrates how each specific marginalized error (on $w$ and $\Omega_M$) varies with $z_{min}$ as we change the error on the cluster priors. Looking at Fig.~\ref{fig:sigma_zmin}, an improvement of a factor of $\sim$2-4 on both $\sigma_M$ and $\sigma_w$ may be achieved with decreased uncertainties (compare solid to dotted lines). 

Currently, the only way to attain uncertainties in the mass parameters of the smallest order in Figures \ref{fig:inv_area} and \ref{fig:sigma_zmin} requires a stacking analysis. For example, see cosmological constraints derived from a weak lensing analysis in Ref. \cite{Rozo2011} as well as stacked phase-space analyses in Ref. \cite{Gifford2017}. Similarly, achieving $\sigma_{\beta}=0.02$  will entail stacking clusters and/or developing some other approach that has not yet been fully investigated. Thus, the dotted line in Fig.~\ref{fig:inv_area} represents not 100 clusters, but 100 cluster ensembles with high-precision mean masses and mean $\beta$'s. Each "cluster ensemble" is built from a number of individual noisy weak lensing cluster profiles and poorly sampled cluster phase spaces. One thing to consider in a future stacked phase spaces analysis is that systematic uncertainties (e.g,. cluster mis-centering) must be accounted for at high precision.

From an observational perspective, it is an interesting question whether one should expend resources on increasing the sample size, or on decreasing the systematic uncertainties. Given a Planck cosmology, to $z=0.8$ we expect to have over 40,000 clusters with $M_{200} > 4\times10^{14}M_{\odot}$ with respect to 200$\times$ the mean density of the Universe \citep{Planck16}. Thus, it seems reasonable to expect that 1000 of these clusters will eventually have both weak-lensing mass estimates and well-sampled radius/velocity phase spaces. Such an effort would require photometry and spectroscopy over about 1000 square degrees of the sky. As an example, the Dark Energy  Spectroscopic Instrument--DESI \footnote{http://desi.lbl.gov/}) is targeting over 1000 square degrees of the Dark Energy Survey sky coverage \citep{DESI}. Likewise, the Prime Focus Camera will provide significant multi-object spectroscopy over more than 1000 square degrees of imaging taken with the Subaru Hyper Suprime-Cam \citep{Takada14}. There is also new PI-based instrumentation, such as the Michigan-Magellan Fiber System (M2FS) on the Magellan observatory, which can be used to specifically and efficiently target clusters with previously measured weak lensing masses \citep{Mateo12}. Thus, it is realistic to expect 1000 clusters with densely sampled phase spaces and redshift errors $\sim 50$ kms$^{-1}$ and weak lensing mass errors of 40\% or less (statistical) in the near future. The technology to collect the needed spectra either exists or is being constructed with the aim to achieve redshift errors on par with existing surveys \citep{Bolton12} and the weak lensing mass errors of 40\% or less  are already being achieved with current imagers \citep{melchior,applegate}. Therefore, in Figs.~\ref{om_w}, \ref{w0wa_joint} and \ref{wOm_joint} we  show the constraints after increasing the fiducial sample size from 100 to 1000 clusters. 

Lastly, it should be obvious that a combination of both more clusters and reduced systematic error would be the optimal solution which yields the tightest constraints. Our analysis in this section is premised on the assumption that both of these options may not be easily available. 

%%%%%%%%%%%%%%%%
%%%%%%%%%%%%%%%%
%%% Discussion %%%
%%%%%%%%%%%%%%%%
%%%%%%%%%%%%%%%%
\section{Discussion}
Other than by reducing statistical errors, increasing the number of clusters, and stacking, we may in principle improve the constraints through a joint likelihood analysis with other cosmological probes. In this section we discuss our constraints and their degeneracies in the context of other probes.

We note that in Fig. \ref{oM_oL} an improvement in the forecasted constraints can be achieved after applying the prior on $h$ from. Ref. \cite{riess2016}. Information on the Hubble parameter breaks numerous degeneracies in our probe. After all, our technique itself is fundamentally based on constraining $qH^2$ (see Sec. \ref{qH2}). However, by including cosmological dependencies on the radial coordinate (Eq.~( \ref{eq:Da})) the probe clearly has some power in constraining $h$ on its own. This is evident in the derivatives shown in Fig. \ref{fig:deriv_four}. If we drop this dependence and use a fixed (in Mpc) radial coordinate, $\Omega_{\Lambda}$, $w_0$, and $w_a$ all become entirely unconstrained. 

\begin{figure}
  \centering
  \includegraphics[width=1\linewidth]{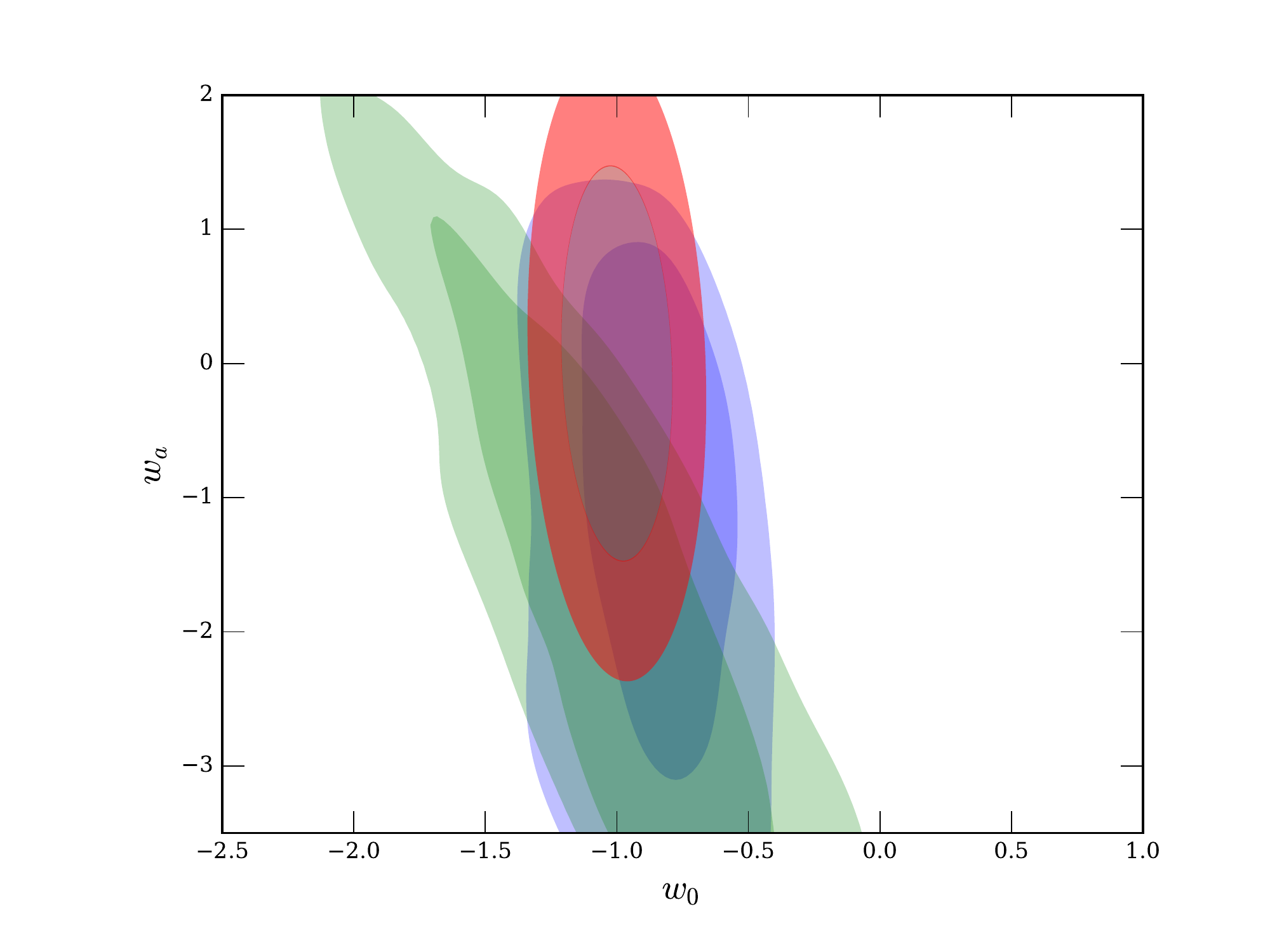}  \caption{68\% and 95\% contours for the dynamic dark energy case using the CPL parametrization of dark energy marginalized over $\Omega_M$ and $h$ as well as the other cluster parameters. The blue contours are reproduced from the latest JLA SNIa constraints of Ref. \cite{JLA}. The green contours are reproduced from the Planck 2015 TT likelihood of Ref. \cite{planck2016}. In red are constraints derived from a sample of $N_{clus} = 1000$ clusters uniformly distributed in the redshift range $ 0 \leq z \leq 0.8$, after applying a conservative $80\%$ mass scatter prior (same as the turquoise contours of Fig.~\ref{w0_wa}). In all cases, no prior assumptions about the Hubble constant are used.}\label{w0wa_joint}
\end{figure}

Besides applying a prior on the Hubble constant, another way we may achieve a tight constraint of the $w_0-w_a$ plane is shown in Fig.~\ref{w0wa_joint}. We show both the 68\% and 95\% confidence constraints with $N_{clus} = 1000$ uniformly distributed in the redshift range $ 0 \leq z \leq 0.8$, after applying a conservative $80\%$ mass scatter prior (in red, same as turquoise contours in Fig.~\ref{w0_wa}) as well as both the JLA SNIa constraints of Ref. \cite{JLA} (in blue) and the 2015 Planck TT likelihood constraints of Ref. \cite{planck2016} (in green). A joint analysis with these probes then seems to have the potential of yielding similar constraints to what a combination of JLA data and CMB currently yields.

\begin{figure}
  \centering
  \includegraphics[width=1\linewidth]{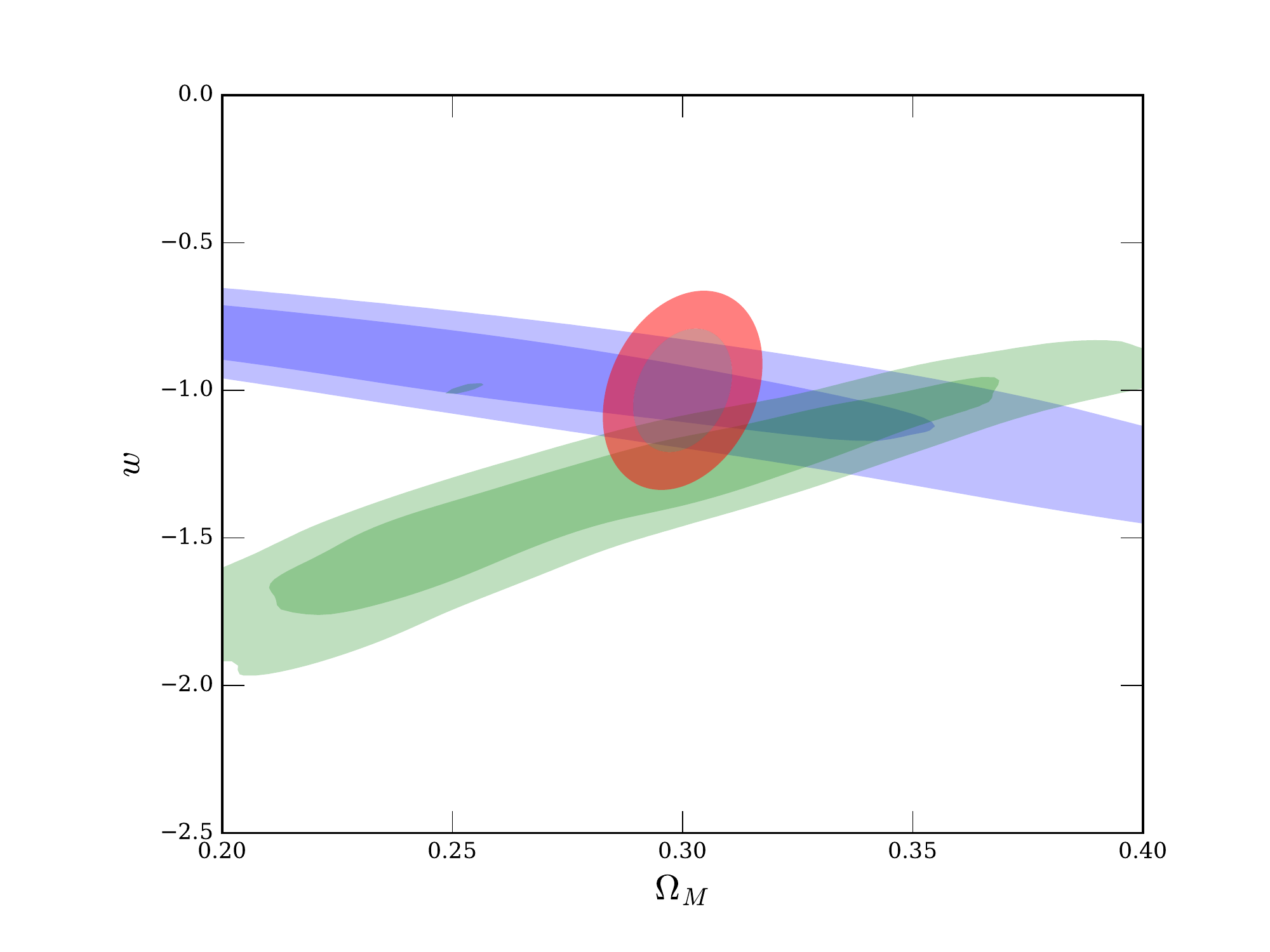}  \caption{68\% and 95\% contours for the flat $w$CDM case. The red contours are derived the same as what is shown in  Fig.~\ref{om_w}. The blue contours are reproduced from the latest JLA SNIa constraints shown in Ref. \cite{JLA}. The green contours are reproduced from the Planck 2015 TT likelihood \cite{planck2016}. In all cases, no prior assumptions about the Hubble constant are used.}\label{wOm_joint}
\end{figure}

Now considering the flat $w$CDM case, in Fig.~\ref{wOm_joint} we over-plot the JLA constraints (in blue), the 2015 Planck TT likelihood of Ref. \cite{planck2016} (in green) as well as re-plot the red contours of Fig.~\ref{om_w}. We find that a joint constraint of these probes alone can yield a joint $ \sigma_{w} \sim 0.1$ and $ \sigma_{\Omega_M} \sim 0.01$  constraint given that they cross through each other nearly perpendicularly. 

We note that the degeneracies in our observable work out in such a way that our probe can act as a powerful cross check of systematics in other probes. As an example, note how our constraints lie perpendicular to the JLA SNIa and CMB constraints in Fig.~\ref{wOm_joint}. In part, this is due to the way degeneracies work with our observable.

As shown in Section II, cosmology in our probe comes in through the quantity $qH^2$ which is both a function of the Hubble parameter and its derivative, $dH(z)/dz$. We emphasize this because our probe in this sense is a true dynamical probe of the expansion history of the universe, similar to the redshift drift. Compare for instance, our constraint degeneracies on the $w_0 - w_a$ plane to those of Ref. \cite{redshiftdrift} (in particular, see Fig. 3).

Beyond these comparisons to other probes, we may ask ourselves if we can justify our observable's sensitivity to cosmology on its own basis. As discussed before, the sensitivity of our probe can be directly inferred from both Fig.~\ref{fig:deriv_four} and Fig.~\ref{fig3}. Recall that on Fig.~\ref{fig3} we plot the fractional difference of the escape velocity profile between the  $\Lambda$CDM model ($w=-1$) and two other dark energy models. At about the virial radius, the fractional difference amounts to $\sim 15\%$. So to a crude first approximation, we need the error budget in the observational parameters to drop below this limit in order to place constraints on $w$. 

As explained before, and as tabulated in Table 1, three sources of observational error are involved: the escape velocity edge error ($\sigma_{v_{esc}}$), the anisotropy parameter uncertainty ($\sigma_{\beta}$) and the error in the inferred Einasto parameters from weak lensing ($\alpha$, $\sigma_{\rho_{-2}}$,$\sigma_{r_{-2}}$). If we focus just on the $\sigma_{v_{esc}}$ we notice that this amounts to a $\sim 15\%$ error on the escape velocity edge. As such, with just one cluster we are on the verge of being able to detect deviations from the $\Lambda$CDM model. Similarly, for our fiducial cluster, the uncertainty in $\sigma_{\beta}=0.5$ amounts to a difference in the escape velocity profile also of $\sim 15 \%$ given that it comes in to our observable through the factor of $1/\sqrt{g(\beta)}$ (see Eq.~(\ref{eq:vesc})). So again, with this systematic uncertainty we are close to being able to detect deviations from $w=-1$. Now let us consider the dominant source of error which comes in through the uncertainty in the inferred Einasto parameters from weak lensing. For 80\% error on the mass, this amounts to an error on the edge of $\sim 40\%$ given that the velocity goes as the square root of the mass. With just 8 clusters we can naively decrease the weak lensing error to ($\sim 15$\%), assuming that it scales as $1/\sqrt{N_{clus}}.$ Of course the above are unrealistic conditions, which is why instead we conduct a detailed Fisher Matrix analysis.  

Another key factor in this probe is that unlike supernovae observations, the cluster data map a projected radial profile which increases the total amount of information per object, thereby further beating down the error. We previously addressed how the binning can affect the final predictions on the cosmological parameters. The key point is that with just a few tens of clusters, this probe becomes sensitive to 15\% deviations in the dark energy equation of state exemplified in Fig.~\ref{fig3}.  While these are only forecasts, as a consistency check, we have compared our Fisher matrix constraints with the analysis of Ref. \cite{starkApJ} which utilized $N_{clus} = 20$ ($ 0 \leq z \leq 0.439 $). We find that our Fisher matrix forecasts are consistent with variations of systematics studied in Ref. \cite{starkApJ}.

%%%%%%%%%%%%%%%%
%%%%%%%%%%%%%%%%
%%% conc %%%
%%%%%%%%%%%%%%%%
%%%%%%%%%%%%%%%%
\section{Conclusions}
We have presented a novel galaxy cluster-based probe of cosmology that has the potential of constraining cosmological parameters to high precision. More specifically, this cosmological probe is based on both the abstract and concrete need to include a cosmological term in the escape velocity profile of galaxy clusters as inferred from their phase spaces. This phase space-inferred escape velocity profile is modeled by cluster-specific parameters (i.e. weak lensing mass profile information and the cluster's anisotropy parameter) as well as cosmological parameters. If the first set of parameters can be independently inferred, then  cosmology can be allowed to vary to fit the observed escape velocity profiles --- thereby constraining cosmological models. 

To assess this probe's observational viability we used the Fisher matrix formalism and carefully considered the aforementioned systematics by marginalizing over the free parameters describing the gravitational potential of each cluster separately. While constraints can be improved if systematic errors in both the weak lensing mass estimates and inferences of the anisotropy parameter are reduced, we note that the the gains are similarly improved by increasing the number of clusters $N_{clus}$. A combination of both of these approaches would of course be optimal. However, we note that even assuming conservative errors, competitive cosmological constraints can still be achieved in the near term. 

It is also important to realize, as mentioned above, that this probe can currently constrain only accelerating cosmologies. In particular, the balance of forces argument with which we derive our theoretical expectation for the escape velocity profile is not valid in a non-accelerating universe. More theoretical work needs to be done to be able to make the theoretical expectation sensitive to cosmology at epochs beyond the transition redshift.

Nonetheless, we have shown that this probe is not only able to yield high precision constraints on cosmological parameters independently of other probes but that it  complements other constraints as well. Furthermore, we emphasize that these constraints can be achieved in both the near and far future. For instance, Fig.~\ref{w0_wa} and Fig.~\ref{oM_oL} only use 100 clusters with 40-80\% weak lensing mass scatter which can easily be achieved in the near term; this is also the case with the black contours constraints of Fig.~\ref{om_w}. Far future constraints ($N_{clus} = 1000 $) are forecasted in Fig.~\ref{om_w} (red contours)  as well as in the future joint constraints of Figs. \ref{w0wa_joint}-\ref{wOm_joint}.

We also note that while throughout this paper we have assumed the general relativistic Friedmann equation, our theoretical expectations can in principle be generalized by re-working the term $qH^2$ to either reflect modified theories of gravity or other alternative parametrizations. 

This work therefore presents a first step in the study of a promising new probe of cosmology. The cluster phase spaces, we demonstrated here, have the power to provide precision measurements of cosmological parameters in an accelerating universe, and thus provide sharp tests of the currently favored theoretical framework.

%%%%%%%%%%%%%%%%
%%%%%%%%%%%%%%%%
%%% Acknowledgements %%%
%%%%%%%%%%%%%%%%
%%%%%%%%%%%%%%%%

\section{acknowledgments}
 AS and CJM are supported by the National Science Foundation under Grant No. 1311820 and 1256260 and the Department of Energy grant DE-SC0013520. DH is supported by NSF grant AST-0807564, DOE grant DE-FG02-95ER40899, and NASA grant NNX16AI41G. The authors would like to thank Eric Linder and the anonymous referee for useful comments. 

%%%%%%%%%%%%%%%%
%%%%%%%%%%%%%%%%
%%% Appendix %%%
%%%%%%%%%%%%%%%%
%%%%%%%%%%%%%%%%

\begin{figure*}
  \centering
  \includegraphics[width=0.8\linewidth]{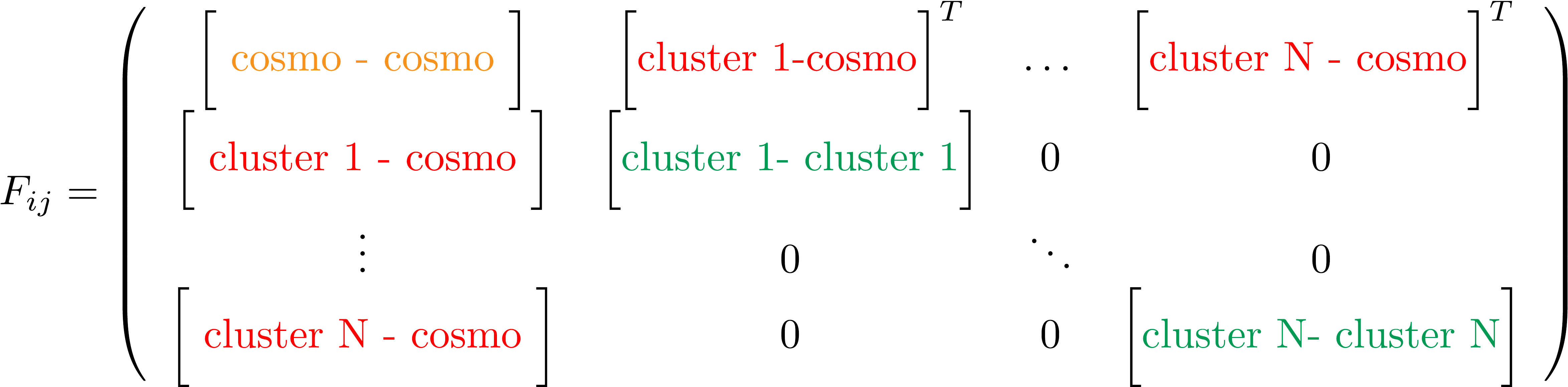}
  \caption{Structure of the $N_{dim} \times N_{dim}$ (see Eq.~(\ref{Ndim})) $F_{ij}$ matrix. The "cosmo-cosmo"  submatrix (orange) has dimensions $N_{cosmo} \times N_{cosmo}$  and contains the information solely of the cosmological parameters and their inverse covariances. The "cosmo-cluster" submatrices  (red) lie along the first row and column of the $F_{ij}$ matrix and are composed of the cross-correlated information between cluster parameters and cosmological parameters. Note that along the first column these submatrices have dimension $N_{clus} \times N_{cosmo} $, and along the first row, the matrices are transposed and therefore have dimensions $N_{cosmo} \times N_{clus}.$ Lastly, the "cluster-cluster" submatrices (green) lie along the diagonal and have dimensions $N_{clus} \times N_{clus}.$ Note that zeros populate the off-diagonal spaces given that there is no correlation between clusters, so that the derivatives cancel out. }\label{Fij}
\end{figure*}

\appendix
\section{$F_{ij}$ matrix structure}
In this section we detail the structure of the $F_{ij}$ matrix. A schematic of the matrix and its submatrices is shown in Fig.~\ref{Fij}. As indicated by the figure, there are three main component submatrices to the $F_{ij}$ matrix: the "cosmo-cosmo" submatrix (orange), the "cluster-cosmo" submatrices (red), and the "cluster-cluster" submatrices (green). The zeros indicate that the "off-diagonal" terms are nil. We describe the components of these submatrices in the following three subsections.

%%%%%%%%%%%%
%%%%%%%%%%%%
\subsection{"cosmo-cosmo" submatrix}
In the top left of the matrix (see Fig.~\ref{Fij}) we have an $N_{cosmo}$ by $N_{cosmo}$ submatrix which is composed exclusively of the derivatives of our observable with respect to cosmological parameters.  Let's consider the flat $w$CDM case, $p_{cosmo} \in \{ \Omega_M, w, h\} $, and take a look at some terms. 
For this case, the first term in this submatrix is then,
\begin{equation} 
F_{00}=  \frac{1}{\sigma_{v_{esc}}^2} \sum_{n,k}    \frac{\partial v_{esc}(z_n,r_k)  }{\partial \Omega_M} \frac{\partial v_{esc}(z_n,r_k)  }{\partial \Omega_M}.
\end{equation}
The off-diagonal term in the first column second row is simply,
\begin{equation} 
F_{01}=  \frac{1}{\sigma_{v_{esc}}^2} \sum_{n,k}    \frac{\partial v_{esc}(z_n,r_k)  }{\partial \Omega_M} \frac{\partial v_{esc}(z_n,r_k)  }{\partial w}. 
\end{equation}
Therefore, we are adding information on some cosmological parameter (or a combination, as in the off diagonal term) both across $n$ clusters and $k$ radial bins. As we detail in the next two sections, this is not the case for
all other elements in the $F_{ij}$ matrix.
%%%%%%%%%%%%
%%%%%%%%%%%%
\subsection{"cosmo-cluster" submatrices}
Now let us take a look at the "cosmo-cluster" submatrices of Fig.~\ref{Fij} (shown in red). Staying in the first row but now looking at the fourth column, we are now looking at the cross information attained from cosmology and cluster parameters. In this case, the anisotropy parameter for cluster 1 ($\beta_1$), the matrix element is,
\begin{multline} 
F_{03}=   \frac{1}{\sigma_{v_{esc}}^2} \sum_{n,k}  \frac{\partial v_{esc}(z_n,r_k)  }{\partial \Omega_M} \frac{\partial v_{esc}(z_n,r_k)}{\partial \beta_1}  \\
=  \frac{1}{\sigma_{v_{esc}}^2} \sum_{k}  \frac{\partial v_{esc}(z_1,r_k)  }{\partial \Omega_M}   \frac{\partial  v_{esc}(z_1,r_k)}{\partial \beta_1}  +  \\ 
\frac{1}{\sigma_{v_{esc}}^2} \sum_{k}  \frac{\partial v_{esc}(z_2,r_k)  }{\partial \Omega_M}   \cancelto{0}{ \frac{\partial  v_{esc}(z_2,r_k)}{\partial \beta_1} }+ \dots 
\end{multline} 

Immediately we notice that the second term of the second sum (i.e. the derivative with respect to $\beta_1$ for the second cluster $z_2$), is nil. Therefore, unlike the "cosmo-cosmo" submatrices, in these submatrices we only sum over the $k$th radial bin of the cluster corresponding to that column. This is the case for subsequent columns and rows (by symmetry). For instance, if $N_{cosmo} = 3$ then, $F_{0j}, F_{1j}, F_{2j}$ where $j = \{ 3, 4, 5, \dots, N_{dim} \}$. Symmetry yields the same for $F_{i0}, F_{i1}, F_{i2}$ where $i = \{ 3, 4, 5, \dots, N_{dim} \}$. The structure is the same for the "cluster-cosmo" submatrices along the first column, where the submatrices are simply transposed, as evoked by the superscript $T$ in Fig.~\ref{Fij}.

%%%%%%%%%%%%
%%%%%%%%%%%%
\subsection{"cluster-cluster" submatrices}
Lastly, let us now take a look at the "cluster-cluster" submatrices of Fig.~\ref{Fij} (green). These submatrices express simply the covariance between cluster parameters for a given cluster. Taking a look at the first element of the first submatrix on the diagonal (the "cluster 1-cluster 1" submatrix of Fig.~\ref{Fij}),
\begin{multline} 
F_{33}=   \frac{1}{\sigma_{v_{esc}}^2} \sum_{n,k}  \frac{\partial v_{esc}(z_n,r_k)  }{\partial \beta_1}   \frac{\partial  v_{esc}(z_n,r_k)}{\partial \beta_1}  \\
=   \frac{1}{\sigma_{v_{esc}}^2} \sum_{k}  \frac{\partial v_{esc}(z_1,r_k)  }{\partial \beta_1}   \frac{\partial  v_{esc}(z_1,r_k)}{\partial \beta_1}  + \\ \frac{1}{\sigma_{v_{esc}}^2} \sum_{k}  \cancelto{0}{\frac{\partial v_{esc}(z_2,r_k)  }{\partial \beta_1}   \frac{\partial  v_{esc}(z_2,r_k)}{\partial \beta_1}} + \dots \\
\end{multline} 
We see that the second term and on will yield 0 given that they are the derivatives of clusters $z_{n \neq1}$ with respect to $\beta_1$. This simply demonstrates that there is no cross correlation between clusters, as expected. Therefore, along the diagonal of $F_{ij}$ we have 4 $\times$ 4 matrices of the various cluster parameters with respect to a given cluster, from cluster 1 to cluster $N_{dim}.$

Let us take a look now at some of the off diagonal submatrices, say between cluster 1 and cluster 2. The first element is,
\begin{multline} 
F_{73}=   \frac{1}{\sigma_{v_{esc}}^2} \sum_{n,k}  \frac{\partial v_{esc}(z_n,r_k)  }{\partial \beta_2}   \frac{\partial  v_{esc}(z_n,r_k)}{\partial \beta_1}  \\
=   \frac{1}{\sigma_{v_{esc}}^2} \sum_{k}  \cancelto{0}{\frac{\partial v_{esc}(z_1,r_k)  }{\partial \beta_2}}  \frac{\partial  v_{esc}(z_1,r_k)}{\partial \beta_1}  + \\ \frac{1}{\sigma_{v_{esc}}^2} \sum_{k}  \frac{\partial v_{esc}(z_2,r_k)  }{\partial \beta_2}   \cancelto{0}{\frac{\partial  v_{esc}(z_2,r_k)}{\partial \beta_1}} + \dots \\
\end{multline} 

Note that the first term in the sum over the radial bins of the first cluster ($n=1$) is nil, and so is the second term of the sum over the radial bins of the second cluster ($n=2$). By induction, all other terms are also nil. Therefore, these off diagonal terms are all zero given that there is no cross correlation between cluster parameters of different clusters. All of these terms are aptly represented by "0"'s in Fig.~ \ref{Fij}.

\section{$F_{prior}$ matrix sub-structure}
In this section we describe the structure of the prior matrix found in Eq.~(\ref{eq:Fmatrix}). In particular, we focus on the structure of the submatrix elements of the $F_{prior}$ matrix in Eq.~(\ref{eq:Fpriorstructure}). The covariance submatrices that lie along the diagonal of Eq.~(\ref{eq:Fpriorstructure}) are given by,

\begin{equation}
C_{cluster} = \begin{bmatrix} \sigma_{\beta}^2 & 0 & 0 & 0 \\    0 & \sigma_{\alpha}^2  & 0 & 0 \\ \ 0 & 0 & \sigma_{r_{-2}}^2 & -0.7 \sigma_{\rho_{-2}} \sigma_{r_{-2}} \\0 & 0 & -0.7 \sigma_{\rho_{-2}} \sigma_{r_{-2}} & \sigma_{\rho_{-2}}^2   \end{bmatrix}.
\end{equation}
Note that the only non-zero terms off the diagonal is the covariance between $r_{-2}$ and $\rho_{-2}$.  Specific values for these matrix elements and the code used to produce the matrices from which we derive the  constraints on cosmological parameters can be found online at \url{https://github.com/alejostark/phase_space_cosmo_fisher}.

%%%%%%%%%%%%
%%%%%%%%%%%%

\section{Weak lensing mass prior and cosmology}

We now consider how uncertainties in the cosmological parameters affect the weak lensing mass uncertainties, which, recall, are in turn featured in our prior information matrix $C_{cluster}$. 

To do this, we carry out a quantitative investigation utilizing the Cluster-Lensing code of Ref. \cite{clusterlensing}. We start out by building a $\Sigma(r)$ surface density shear profile for one fiducial cluster with $M_{200} = 4\times 10^{14} M_{\odot}$. We first create this profile assuming  fixed, fiducial values of the cosmological parameters. We assume the profile has Gaussian errors of such size to ensure that we recover a 20\% statistical error on the cluster mass after performing a simple $\chi^2$ analysis. 

We then allow the cosmological parameters to vary, and conduct a Markov Chain Monte Carlo (MCMC) analysis with $emcee$ to sample the posterior distribution and examine the likelihood of the inferred mass $M_{200}$ \cite{emcee}. The likelihood model is given by

\begin{equation}
\ln \mathcal{L}(\Sigma |r_k, z, \Theta) = -\frac{1}{2} \sum_{k} \frac{\bigg(\Sigma(r_k,z,\Theta_{fid}) - \Sigma(r_k,z,\Theta)\bigg)^2}{\sigma_{\Sigma}^2}.
\end{equation}

Assuming the flat $w$CDM model, our parameter set is  given by $\Theta = \{ \Omega_M , w , h, M_{200} \}$. Note that the cluster profile information is reduced to a single parameter, $M_{200}$, because the Cluster-Lensing code uses a mass-concentration relation to create the $\Sigma(r)$ profile \cite{clusterlensing}.  For a single mock $\Sigma(r)$ profile, after marginalizing over all other parameters we find that the total  error in the cluster mass scale increases from 20\% to 40\%. That is, if cosmological parameters are allowed to vary, the weak lensing mass error increases by a factor of two.  We confirm this using statistical errors of 5\% and also 40\%.

\begin{figure}
  \centering
  \includegraphics[width=1\linewidth]{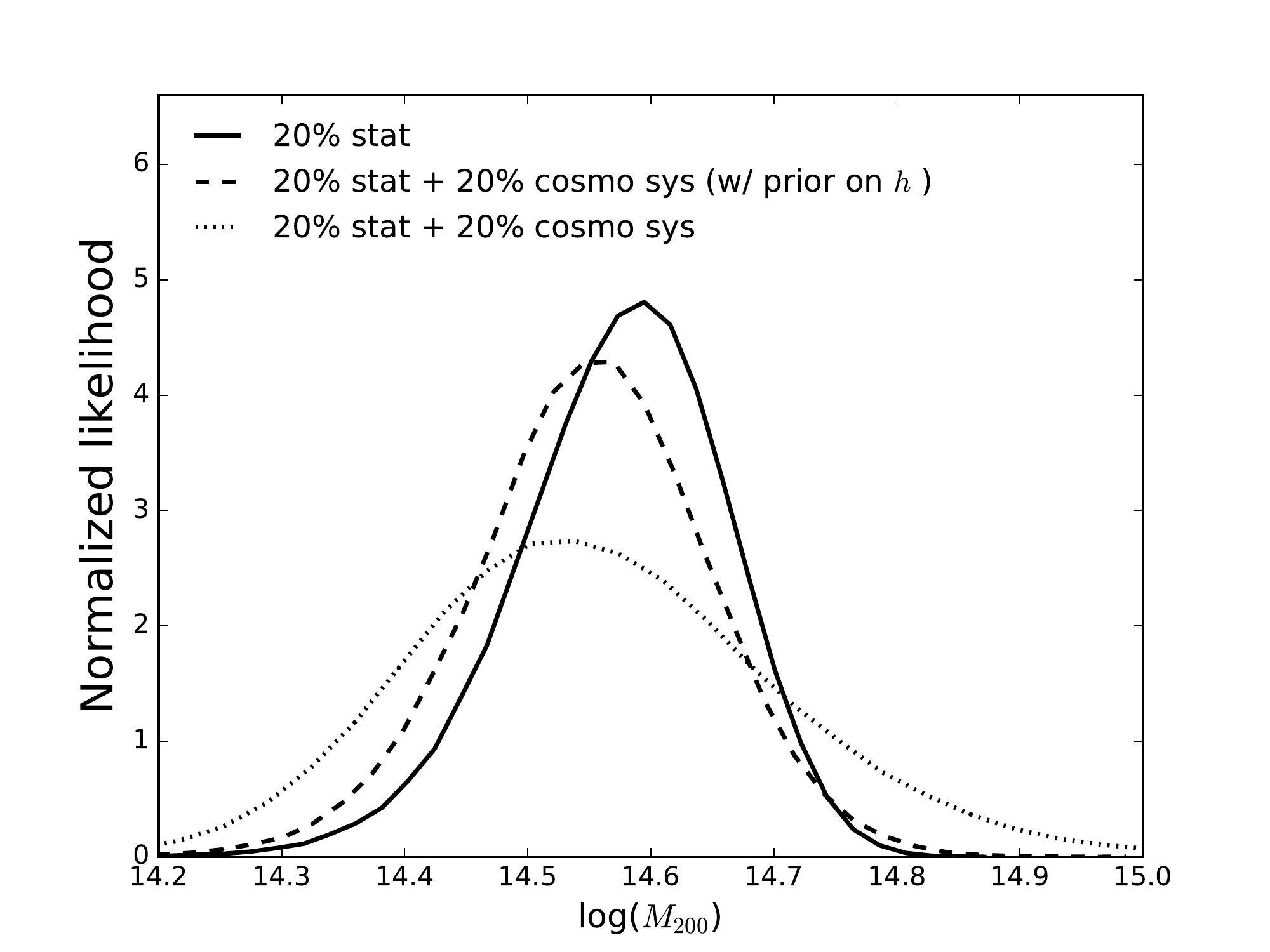}
  \caption{Marginalized likelihood of the inferred mass of our fiducial cluster ($M_{200} = 4\times 10^{14} M_{\odot}$) from the weak lensing surface density shear profile $\Sigma(r)$. The total uncertainty in the inferred mass increases by a factor of $\sim 2$ if no prior on the cosmological parameters is introduced; however, a reasonable prior on the Hubble parameter $h$ from Ref. \cite{riess2016} recovers most of the lost information on $M_{200}$.}\label{logm200}
\end{figure}

We show the marginalized posterior likelihood in Fig.~\ref{logm200}. 
Three different prior likelihoods are shown: a strong prior, basically fixing cosmology (represented by the solid black line), a prior only on $h$ (dashed line), and no prior (dotted line). Note how the 20\% systematic error likelihood is broadened if cosmological parameters are allowed to vary, but that we can almost totally reduce the cosmological systematic error simply by applying a reasonable prior on the Hubble parameter of  Ref. \cite{riess2016}  ($\sigma_h = 0.0174$). We tabulate the resulting uncertainties in the Einasto cluster parameters in Table 1. As stated in the previous subsection, the code used to produce these results can be found online. Lastly, we note that since the bias in the recovered mass (dotted and dashed) is small, and we do not factor it into our analysis.

%%%%%%%%%%%%%%%%%%%%%%%%%%
%%%%%%%%%%%%%%%%%%%%%%%%%%
%%%%%%%%%%%%%%%%%%%%%%%%%%

\bibliographystyle{apsrev4-1}
\bibliography{stark}

\end{document}